\documentclass[journal,draftclsnofoot,onecolumn,12pt]{IEEEtran}
\usepackage{multirow}
\usepackage{array}
\newcolumntype{P}[1]{>{\centering\arraybackslash}p{#1}}
\newcolumntype{M}[1]{>{\centering\arraybackslash}m{#1}}
\usepackage{amsthm,amssymb,amsmath,graphicx,multirow,color,amsfonts}%
\usepackage[update,prepend]{epstopdf}
\usepackage{multirow}
\usepackage[latin1]{inputenc}
\usepackage{tikz}
\usepackage{bbm} 
\usepackage{pdfpages}
\usepackage{multirow}
\usepackage{subfig}
\usepackage{comment}
 \usepackage{graphicx}
\usepackage{multicol}

\usepackage{cite}
\usepackage[justification=centering]{caption}
\usepackage{textcomp}
\usepackage{psfrag}
\usepackage{arydshln}
\usepackage{url}
\usepackage{soul}
\usepackage{graphicx,color}
\usepackage[nolist]{acronym}
\usepackage{algorithm,algorithmic} 

\usepackage{mathtools,lipsum}
\usepackage{cuted}
\usepackage{amsmath}
\newtheorem{proposition}{Proposition} 
\usepackage{mathrsfs}
\usepackage{color}

\usepackage[capitalise]{cleveref}
\Crefname{equation}{Eq.\!}{Eqs.\!}
\Crefname{figure}{Fig.\!}{Figs.\!}
\Crefname{tabular}{Tab.\!}{Tabs.\!}
\Crefname{section}{Section\!}{Sections.\!}

\def\nb0{{\mathbf{0}}}
\def\nb1{{\mathbf{1}}}

\newtheorem{lemma}{Lemma}

\newtheorem{definition}{Definition}

\newtheorem{theorem}{Theorem}

\newtheorem{remark}{Remark}

\begin{document}
\graphicspath{{./Figures/}}
	\begin{acronym}

\acro{5G-NR}{5G New Radio}
\acro{3GPP}{3rd Generation Partnership Project}
\acro{ABS}{aerial base station}
\acro{AC}{address coding}
\acro{ACF}{autocorrelation function}
\acro{ACR}{autocorrelation receiver}
\acro{ADC}{analog-to-digital converter}
\acrodef{aic}[AIC]{Analog-to-Information Converter}     
\acro{AIC}[AIC]{Akaike information criterion}
\acro{aric}[ARIC]{asymmetric restricted isometry constant}
\acro{arip}[ARIP]{asymmetric restricted isometry property}

\acro{ARQ}{Automatic Repeat Request}
\acro{AUB}{asymptotic union bound}
\acrodef{awgn}[AWGN]{Additive White Gaussian Noise}     
\acro{AWGN}{additive white Gaussian noise}

\acro{APSK}[PSK]{asymmetric PSK} 

\acro{waric}[AWRICs]{asymmetric weak restricted isometry constants}
\acro{warip}[AWRIP]{asymmetric weak restricted isometry property}
\acro{BCH}{Bose, Chaudhuri, and Hocquenghem}        
\acro{BCHC}[BCHSC]{BCH based source coding}
\acro{BEP}{bit error probability}
\acro{BFC}{block fading channel}
\acro{BG}[BG]{Bernoulli-Gaussian}
\acro{BGG}{Bernoulli-Generalized Gaussian}
\acro{BPAM}{binary pulse amplitude modulation}
\acro{BPDN}{Basis Pursuit Denoising}
\acro{BPPM}{binary pulse position modulation}
\acro{BPSK}{Binary Phase Shift Keying}
\acro{BPZF}{bandpass zonal filter}
\acro{BSC}{binary symmetric channels}              
\acro{BU}[BU]{Bernoulli-uniform}
\acro{BER}{bit error rate}
\acro{BS}{base station}
\acro{BW}{BandWidth}
\acro{BLLL}{ binary log-linear learning }

\acro{CP}{Cyclic Prefix}
\acrodef{cdf}[CDF]{cumulative distribution function}   
\acro{CDF}{Cumulative Distribution Function}
\acrodef{c.d.f.}[CDF]{cumulative distribution function}
\acro{CCDF}{complementary cumulative distribution function}
\acrodef{ccdf}[CCDF]{complementary CDF}               
\acrodef{c.c.d.f.}[CCDF]{complementary cumulative distribution function}
\acro{CD}{cooperative diversity}

\acro{CDMA}{Code Division Multiple Access}
\acro{ch.f.}{characteristic function}
\acro{CIR}{channel impulse response}
\acro{cosamp}[CoSaMP]{compressive sampling matching pursuit}
\acro{CR}{cognitive radio}
\acro{cs}[CS]{compressed sensing}                   
\acrodef{cscapital}[CS]{Compressed sensing} 
\acrodef{CS}[CS]{compressed sensing}
\acro{CSI}{channel state information}
\acro{CCSDS}{consultative committee for space data systems}
\acro{CC}{convolutional coding}
\acro{Covid19}[COVID-19]{Coronavirus disease}

\acro{DAA}{detect and avoid}
\acro{DAB}{digital audio broadcasting}
\acro{DCT}{discrete cosine transform}
\acro{dft}[DFT]{discrete Fourier transform}
\acro{DR}{distortion-rate}
\acro{DS}{direct sequence}
\acro{DS-SS}{direct-sequence spread-spectrum}
\acro{DTR}{differential transmitted-reference}
\acro{DVB-H}{digital video broadcasting\,--\,handheld}
\acro{DVB-T}{digital video broadcasting\,--\,terrestrial}
\acro{DL}{DownLink}
\acro{DSSS}{Direct Sequence Spread Spectrum}
\acro{DFT-s-OFDM}{Discrete Fourier Transform-spread-Orthogonal Frequency Division Multiplexing}
\acro{DAS}{Distributed Antenna System}
\acro{DNA}{DeoxyriboNucleic Acid}

\acro{EC}{European Commission}
\acro{EED}[EED]{exact eigenvalues distribution}
\acro{EIRP}{Equivalent Isotropically Radiated Power}
\acro{ELP}{equivalent low-pass}
\acro{eMBB}{Enhanced Mobile Broadband}
\acro{EMF}{ElectroMagnetic Field}
\acro{EU}{European union}
\acro{EI}{Exposure Index}
\acro{eICIC}{enhanced Inter-Cell Interference Coordination}

\acro{FC}[FC]{fusion center}
\acro{FCC}{Federal Communications Commission}
\acro{FEC}{forward error correction}
\acro{FFT}{fast Fourier transform}
\acro{FH}{frequency-hopping}
\acro{FH-SS}{frequency-hopping spread-spectrum}
\acrodef{FS}{Frame synchronization}
\acro{FSsmall}[FS]{frame synchronization}  
\acro{FDMA}{Frequency Division Multiple Access}

\acro{GA}{Gaussian approximation}
\acro{GF}{Galois field }
\acro{GG}{Generalized-Gaussian}
\acro{GIC}[GIC]{generalized information criterion}
\acro{GLRT}{generalized likelihood ratio test}
\acro{GPS}{Global Positioning System}
\acro{GMSK}{Gaussian Minimum Shift Keying}
\acro{GSMA}{Global System for Mobile communications Association}
\acro{GS}{ground station}
\acro{GMG}{ Grid-connected MicroGeneration}

\acro{HAP}{high altitude platform}
\acro{HetNet}{Heterogeneous network}

\acro{IDR}{information distortion-rate}
\acro{IFFT}{inverse fast Fourier transform}
\acro{iht}[IHT]{iterative hard thresholding}
\acro{i.i.d.}{independent, identically distributed}
\acro{IoT}{Internet of Things}                      
\acro{IR}{impulse radio}
\acro{lric}[LRIC]{lower restricted isometry constant}
\acro{lrict}[LRICt]{lower restricted isometry constant threshold}
\acro{ISI}{intersymbol interference}
\acro{ITU}{International Telecommunication Union}
\acro{ICNIRP}{International Commission on Non-Ionizing Radiation Protection}
\acro{IEEE}{Institute of Electrical and Electronics Engineers}
\acro{ICES}{IEEE international committee on electromagnetic safety}
\acro{IEC}{International Electrotechnical Commission}
\acro{IARC}{International Agency on Research on Cancer}
\acro{IS-95}{Interim Standard 95}

\acro{KPI}{Key Performance Indicator}

\acro{LEO}{low earth orbit}
\acro{LF}{likelihood function}
\acro{LLF}{log-likelihood function}
\acro{LLR}{log-likelihood ratio}
\acro{LLRT}{log-likelihood ratio test}
\acro{LoS}{Line-of-Sight}
\acro{LRT}{likelihood ratio test}
\acro{wlric}[LWRIC]{lower weak restricted isometry constant}
\acro{wlrict}[LWRICt]{LWRIC threshold}
\acro{LPWAN}{Low Power Wide Area Network}
\acro{LoRaWAN}{Low power long Range Wide Area Network}
\acro{NLoS}{Non-Line-of-Sight}
\acro{LiFi}[Li-Fi]{light-fidelity}
 \acro{LED}{light emitting diode}
 \acro{LABS}{LoS transmission with each ABS}
 \acro{NLABS}{NLoS transmission with each ABS}

\acro{MB}{multiband}
\acro{MC}{macro cell}
\acro{MDS}{mixed distributed source}
\acro{MF}{matched filter}
\acro{m.g.f.}{moment generating function}
\acro{MI}{mutual information}
\acro{MIMO}{Multiple-Input Multiple-Output}
\acro{MISO}{multiple-input single-output}
\acrodef{maxs}[MJSO]{maximum joint support cardinality}                       
\acro{ML}[ML]{maximum likelihood}
\acro{MMSE}{minimum mean-square error}
\acro{MMV}{multiple measurement vectors}
\acrodef{MOS}{model order selection}
\acro{M-PSK}[${M}$-PSK]{$M$-ary phase shift keying}                       
\acro{M-APSK}[${M}$-PSK]{$M$-ary asymmetric PSK} 
\acro{MP}{ multi-period}
\acro{MINLP}{mixed integer non-linear programming}

\acro{M-QAM}[$M$-QAM]{$M$-ary quadrature amplitude modulation}
\acro{MRC}{maximal ratio combiner}                  
\acro{maxs}[MSO]{maximum sparsity order}                                      
\acro{M2M}{Machine-to-Machine}                                                
\acro{MUI}{multi-user interference}
\acro{mMTC}{massive Machine Type Communications}      
\acro{mm-Wave}{millimeter-wave}
\acro{MP}{mobile phone}
\acro{MPE}{maximum permissible exposure}
\acro{MAC}{media access control}
\acro{NB}{narrowband}
\acro{NBI}{narrowband interference}
\acro{NLA}{nonlinear sparse approximation}
\acro{NLOS}{Non-Line of Sight}
\acro{NTIA}{National Telecommunications and Information Administration}
\acro{NTP}{National Toxicology Program}
\acro{NHS}{National Health Service}

\acro{LOS}{Line of Sight}

\acro{OC}{optimum combining}                             
\acro{OC}{optimum combining}
\acro{ODE}{operational distortion-energy}
\acro{ODR}{operational distortion-rate}
\acro{OFDM}{Orthogonal Frequency-Division Multiplexing}
\acro{omp}[OMP]{orthogonal matching pursuit}
\acro{OSMP}[OSMP]{orthogonal subspace matching pursuit}
\acro{OQAM}{offset quadrature amplitude modulation}
\acro{OQPSK}{offset QPSK}
\acro{OFDMA}{Orthogonal Frequency-division Multiple Access}
\acro{OPEX}{Operating Expenditures}
\acro{OQPSK/PM}{OQPSK with phase modulation}

\acro{PAM}{pulse amplitude modulation}
\acro{PAR}{peak-to-average ratio}
\acrodef{pdf}[PDF]{probability density function}                      
\acro{PDF}{probability density function}
\acrodef{p.d.f.}[PDF]{probability distribution function}
\acro{PDP}{power dispersion profile}
\acro{PMF}{probability mass function}                             
\acrodef{p.m.f.}[PMF]{probability mass function}
\acro{PN}{pseudo-noise}
\acro{PPM}{pulse position modulation}
\acro{PRake}{Partial Rake}
\acro{PSD}{power spectral density}
\acro{PSEP}{pairwise synchronization error probability}
\acro{PSK}{phase shift keying}
\acro{PD}{power density}
\acro{8-PSK}[$8$-PSK]{$8$-phase shift keying}
\acro{PPP}{Poisson point process}
\acro{PCP}{Poisson cluster process}
 
\acro{FSK}{Frequency Shift Keying}

\acro{QAM}{Quadrature Amplitude Modulation}
\acro{QPSK}{Quadrature Phase Shift Keying}
\acro{OQPSK/PM}{OQPSK with phase modulator }

\acro{RD}[RD]{raw data}
\acro{RDL}{"random data limit"}
\acro{ric}[RIC]{restricted isometry constant}
\acro{rict}[RICt]{restricted isometry constant threshold}
\acro{rip}[RIP]{restricted isometry property}
\acro{ROC}{receiver operating characteristic}
\acro{rq}[RQ]{Raleigh quotient}
\acro{RS}[RS]{Reed-Solomon}
\acro{RSC}[RSSC]{RS based source coding}
\acro{r.v.}{random variable}                               
\acro{R.V.}{random vector}
\acro{RMS}{root mean square}
\acro{RFR}{radiofrequency radiation}
\acro{RIS}{Reconfigurable Intelligent Surface}
\acro{RNA}{RiboNucleic Acid}
\acro{RRM}{Radio Resource Management}
\acro{RUE}{reference user equipments}
\acro{RAT}{radio access technology}
\acro{RB}{resource block}

\acro{SA}[SA-Music]{subspace-augmented MUSIC with OSMP}
\acro{SC}{small cell}
\acro{SCBSES}[SCBSES]{Source Compression Based Syndrome Encoding Scheme}
\acro{SCM}{sample covariance matrix}
\acro{SEP}{symbol error probability}
\acro{SG}[SG]{sparse-land Gaussian model}
\acro{SIMO}{single-input multiple-output}
\acro{SINR}{signal-to-interference plus noise ratio}
\acro{SIR}{signal-to-interference ratio}
\acro{SISO}{Single-Input Single-Output}
\acro{SMV}{single measurement vector}
\acro{SNR}[\textrm{SNR}]{signal-to-noise ratio} 
\acro{sp}[SP]{subspace pursuit}
\acro{SS}{spread spectrum}
\acro{SW}{sync word}
\acro{SAR}{specific absorption rate}
\acro{SSB}{synchronization signal block}
\acro{SR}{shrink and realign}

\acro{tUAV}{tethered Unmanned Aerial Vehicle}
\acro{TBS}{terrestrial base station}

\acro{uUAV}{untethered Unmanned Aerial Vehicle}
\acro{PDF}{probability density functions}

\acro{PL}{path-loss}

\acro{TH}{time-hopping}
\acro{ToA}{time-of-arrival}
\acro{TR}{transmitted-reference}
\acro{TW}{Tracy-Widom}
\acro{TWDT}{TW Distribution Tail}
\acro{TCM}{trellis coded modulation}
\acro{TDD}{Time-Division Duplexing}
\acro{TDMA}{Time Division Multiple Access}
\acro{Tx}{average transmit}

\acro{UAV}{Unmanned Aerial Vehicle}
\acro{uric}[URIC]{upper restricted isometry constant}
\acro{urict}[URICt]{upper restricted isometry constant threshold}
\acro{UWB}{ultrawide band}
\acro{UWBcap}[UWB]{Ultrawide band}   
\acro{URLLC}{Ultra Reliable Low Latency Communications}
         
\acro{wuric}[UWRIC]{upper weak restricted isometry constant}
\acro{wurict}[UWRICt]{UWRIC threshold}                
\acro{UE}{User Equipment}
\acro{UL}{UpLink}

\acro{WiM}[WiM]{weigh-in-motion}
\acro{WLAN}{wireless local area network}
\acro{wm}[WM]{Wishart matrix}                               
\acroplural{wm}[WM]{Wishart matrices}
\acro{WMAN}{wireless metropolitan area network}
\acro{WPAN}{wireless personal area network}
\acro{wric}[WRIC]{weak restricted isometry constant}
\acro{wrict}[WRICt]{weak restricted isometry constant thresholds}
\acro{wrip}[WRIP]{weak restricted isometry property}
\acro{WSN}{wireless sensor network}                        
\acro{WSS}{Wide-Sense Stationary}
\acro{WHO}{World Health Organization}
\acro{Wi-Fi}{Wireless Fidelity}

\acro{sss}[SpaSoSEnc]{sparse source syndrome encoding}

\acro{VLC}{Visible Light Communication}
\acro{VPN}{Virtual Private Network} 
\acro{RF}{Radio Frequency}
\acro{FSO}{Free Space Optics}
\acro{IoST}{Internet of Space Things}

\acro{GSM}{Global System for Mobile Communications}
\acro{2G}{Second-generation cellular network}
\acro{3G}{Third-generation cellular network}
\acro{4G}{Fourth-generation cellular network}
\acro{5G}{Fifth-generation cellular network}	
\acro{gNB}{next-generation Node-B Base Station}
\acro{NR}{New Radio}
\acro{UMTS}{Universal Mobile Telecommunications Service}
\acro{LTE}{Long Term Evolution}

\acro{QoS}{Quality of Service}
\end{acronym}
	
\newcommand{\SAR} {\mathrm{SAR}}
\newcommand{\WBSAR} {\mathrm{SAR}_{\mathsf{WB}}}
\newcommand{\gSAR} {\mathrm{SAR}_{10\si{\gram}}}
\newcommand{\Sab} {S_{\mathsf{ab}}}
\newcommand{\Eavg} {E_{\mathsf{avg}}}
\newcommand{\ft}{f_{\textsf{th}}}
\newcommand{\alphatf}{\alpha_{24}}

\title{
Reliability Analysis of Multi-hop Routing in Multi-tier LEO Satellite Networks
}
\author{
Ruibo Wang, Mustafa A. Kishk, {\em Member, IEEE} and Mohamed-Slim Alouini, {\em Fellow, IEEE}
\thanks{Ruibo Wang and Mohamed-Slim Alouini are with King Abdullah University of Science and Technology (KAUST), CEMSE division, Thuwal 23955-6900, Saudi Arabia. Mustafa A. Kishk is with the Department of Electronic Engineering, National University of Ireland, Maynooth, W23 F2H6, Ireland. (e-mail: ruibo.wang@kaust.edu.sa; mustafa.kishk@mu.ie;
slim.alouini@kaust.edu.sa). 
}
\vspace{-4mm}
}
\maketitle

\begin{abstract}
This article studies the reliability of multi-hop routing in a multi-tier hybrid satellite-terrestrial relay network (HSTRN). We evaluate the reliability of multi-hop routing by introducing interruption probability, which is the probability that no relay device (ground gateway or satellite) is available during a hop. The single-hop interruption probability is derived and extended to the multi-hop interruption probability using a stochastic geometry-based approach. Since the interruption probability in HSTRN highly depends on the priority of selecting communication devices at different tiers, we propose three priority strategies: (i) stationary optimal priority strategy, (ii) single-hop interruption probability inspired strategy, and (iii) density inspired strategy. Among them, the interruption probability under the stationary optimal priority strategy can approach the ideal lower bound. However, when analyzing an HSTRN with a large number of tiers, the stationary optimal priority strategy is computationally expensive. The single-hop interruption probability inspired strategy is expected to be a low-complexity but less reliable alternative to the stationary optimal priority strategy. In numerical results, we study the complementarity between terrestrial devices and satellites. Furthermore, analytical results for reliability are also applicable to the analysis of satellite availability, coverage probability, and ultra-reliable and low latency communications (URLLC) rate. Finally, we extend our original routing strategy into a multi-flow one with dynamic priority strategy.
\end{abstract}


\section{Introduction}
In recent years, due to the explosive growth of the number of low Earth orbit (LEO) satellites, the satellite network is expected to play a pivotal role in next-generation wireless networks \cite{yue2022security}. Satellite-to-satellite communication follows the free space propagation model, and its signal attenuation is much smaller than that of terrestrial communication \cite{yaacoub2020key}. Combined with their low production and launch cost \cite{9568932}, relatively low latency, and unique ability to provide seamless global coverage \cite{kodheli2020satellite}, satellite networks have irreplaceable advantages in ultra long-distance communications \cite{chaudhry2020free}. With the establishment of mega-constellations, satellites will not only help expand the coverage of terrestrial networks, but will also enable independent routing. The long-distance routing scheme with LEO satellite as the primary communication carrier and ground gateway as the relay has gradually become a reality \cite{zhu2021integrated,zhang2021stochastic}. Routing reliability is an urgent and fundamental problem in such an HSTRN.

\par
To quantify the reliability of HSTRN in real-time communication, we define interruption probability, which is the probability that no available relay device can be found (satellite or ground gateway) for each hop of a multi-hop routing \cite{wang2022stochastic}. However, the special topology of satellite routing determines that reliability analysis is different from previous research. The particularity of satellite routing can be summarized in the following three points. Firstly, the link between satellites may be blocked by the Earth, resulting in no direct path \cite{Al-3}. Secondly, many satellite constellations are not large enough for independent routing \cite{yue2022security}, and a multi-constellation cooperative routing is more realistic \cite{tang2018multipath}. In addition, several satellite companies, such as OneWeb and Kuiper, deploy satellites at different altitudes \cite{9568932,petit2021assessment}. Therefore, it is meaningful to analyze the interruption probability of a multi-tier HSTRN. Finally, the routing can end within a few hops since the communication distance of a single hop is not much shorter than the total routing distance. Therefore, assuming each hop of the route is independent is not reasonable under a small number of hops. In summary, the limitation of single-hop distance, the requirement of cross-tier spherical routing, and the correlation between hops make interruption probability analysis challenging.

\subsection{Related Works}
Although reliability has been studied in both static (such as sensor networks) and dynamic  networks \cite{lou2021green} (such as unmanned aerial vehicles (UAV) networks), we will pay more attention to the reliability research of satellite networks because these networks are quite different from satellite networks in topology. So far, some studies have established the foundation and brought inspiration for the reliability of LEO satellite routing. Some existing pieces of literature study the existence of reachable links between two satellites by using the graph theory model \cite{shen2020dynamic_2, geng2021agent_3, 9348676_1}. This kind of modeling requires designing a deterministic network topology \cite{9348676_1,shen2020dynamic_2}, or referring to other existing network models \cite{geng2021agent_3,knight2011internet}. Since LEO satellites are not geosynchronous, the reliability analysis based on deterministic topology is not applicable. In addition, graph theory-based models require pre-set starting and ending positions \cite{9348676_1,shen2020dynamic_2}. In addition to the graph theory model, there are also studies on routing reliability estimation based on random algorithms, such as the ant colony algorithm \cite{zhao2021multi_4} and particle swarm optimization algorithm \cite{8068282_5}. Although the proposed methods can be applied to routing in a dynamic system, these studies can only provide numerical simulation results and are difficult to be supported by theoretical analysis. Sometimes, deriving analytical expressions about reliability for satellite constellations are more attractive and convincing than algorithm design and numerical simulation.

\par
Since the above methods are unsuitable for the reliability analysis of a dynamic HSTRN, it is crucial to introduce an effective mathematical tool. Among numerous modeling tools, the model based on stochastic geometry is undoubtedly one of the most suitable models for dynamic network topology analysis \cite{wang2022ultra}. Based on the stochastic geometry analytical framework, several articles have studied the routing on two-dimensional planes \cite{dhillon2015wireless,farooq2015stochastic,sasaki2017energy,routingimportant,richter2018optimal}, which can be extended to reliability analysis for sensor and UAV network, and coverage probability of satellite network \cite{wang2022evaluating,Al-1,Al-2, talgat2020nearest,talgat2020stochastic,ok-1,ok-2,al2021modeling}, which lays a solid foundation for routing in spherical HSTRN network. The modeling of satellite constellations under the stochastic geometry framework is worth emphasizing. Spherical binomial point process (BPP) \cite{talgat2020nearest,ok-1} and spherical Poisson point process (PPP) \cite{Al-1,Al-2} are the most common models for satellite constellations. Although there are some differences with the actual constellation model, BPP and PPP have strong tractability and are therefore widely applied by scholars. Furthermore, BPP models are proved to have little difference in performance analysis with real deterministic constellations \cite{ok-1}. With the increasing size of constellations, the difference between BPP and constellation in network topology will be smaller \cite{wang2022evaluating}. In addition, satellite constellations at different altitudes are usually operated by different companies, targeted at different user groups, and have different commercial applications. Therefore, the authors in \cite{talgat2020nearest, talgat2020stochastic} model a multi-tier satellite network where each tier is an independent homogeneous BPP.

\par
Some of the above studies provide direct references for the routing reliability analysis in this article.  The introduction of BPP connects the probability of finding relay devices in a given region with the area of the region, which provides an effective method for the calculation of interruption probability \cite{wang2022conditional}. In the case of no available satellites in line-of-sight, \cite{ok-1} gives the probability of no available satellite, which provides references for the analysis of interruption probability. The authors in \cite{sasaki2017energy} and \cite{routingimportant} ensure the routing can always follow the established direction by limiting routing to a sector. The coverage analysis of satellite networks under the limitation of maximum reliable communication distance is provided in \cite{Al-1,Al-2}. The routing strategy when only local information is available has been studied \cite{richter2018optimal}, which fits the LEO satellite communication well. Finally, the reliability of single-tier satellite network routing is also discussed in the author's previous research \cite{wang2022stochastic}. However, this article's reliability analysis is only based on a specific algorithm and does not provide an analytical expression of the interruption probability. Therefore, this article will provide the analysis results of multi-hop routing reliability for multi-tier HSTRN.

\subsection{Contributions}
To our best knowledge, this is the first study about reliability analysis in HSTRN using stochastic geometry tools. The contributions can be summarized as follows:
\begin{itemize}
    \item We propose a routing scheme suitable for the stochastic geometry analysis framework. As a measure of reliability, an analytical expression for the multi-hop interruption probability is given and its uniqueness is proved. The analytical expressions of its intermediate results: single-hop interruption probability, the average number of hops before the interruption, and the average number of hops required for successful transmission, are also estimated.
    \item Through a case study of a three-tier  HSTRN, the accuracy of the above analytical results is verified. How to apply the analytical results to this HSTRN is explained in detail. 
    \item As an important factor influencing the interruption probability, three priority strategies involving stationary optimal priority strategy (proposed), single-hop interruption probability inspired strategy, and density inspired strategy are discussed. They sacrifice reliability for computational complexity gains to different degrees.
    \item In addition to the basic simulation results in the above case study, we present more attractive numerical results as further extensions. The performance of other widely studied metrics, including satellite availability, coverage probability, and URLLC rate, are provided. Then, the interruption probability in multi-flow routing scheme and dynamic priority strategy are presented, respectively. 
\end{itemize}

\section{Single-hop Reliability Analysis} \label{Single}
In this section, we first describe the spatial model and communication technique of LEO satellites and gateways. Then, the analytical expression of the tier-to-tier interruption probability for a single-hop is provided. Finally, we derive an analytical expression for the total interruption probability for a single-hop.


\begin{table*}[]
\centering
\caption{Summary of notation.}
\begin{tabular}{|M{3.2cm}|M{11cm}|}
\hline
\textbf{Notation}                                            & \textbf{Description}           \\ \hline  \hline       $R_k$; $h_k$; $N_k$            &Radius of tier $k$; height of tier $k$; the number of devices in tier $k$ \\ \hline
$\theta_m$       & Dome angle of the ground transmitter and receiver         \\ \hline
$\theta_{r}$; $\theta_{s}$; $\theta_{i,j}$; $d_{th}$ & Maximum direction angle; minimum dome angle; maximum dome angle; maximum reliable communication distance      \\ \hline
$P_{i,j}^{I}$;  $P_{i}^{S}$ ; $\overline{P}^S$ &
Interruption probability from tier $i$ to tier $j$; single-hop interruption probability from tier $i$; weighted single-hop interruption probability \\  \hline   $\widetilde{P}^M$; $P^C$ &
Approximate multi-hop interruption probability; Cumulative interruption probability \\ \hline
$s$;   $s^*$; $\mathcal{S}$      & Priority strategy; stationary optimal priority strategy; the set of all priority strategies  \\ \hline
$T^{(1)}$, $\widetilde{T}^{(2)}$, $\widehat{T}^{(3)}$; \ \ \ $\mathcal{T}^{(1)}$, $\widetilde{\mathcal{T}}^{(2)}$, $\widehat{\mathcal{T}}^{(3)}$ &
  TPM and its variations; Operators used to obtain TPMs\\ \hline
$\mu_i$; $N_h$, $\overline{\theta}_o$  & Average number of hops before interruption from tier $i$; average number of hops required for successful transmission, average dome angle per step forward \\ \hline
$\mathcal{K}$    &  Index set of reachable tiers  \\ \hline
\end{tabular} 
\end{table*}

\subsection{System Model}
Consider a scenario where multiple constellations work with ground networks to complete routing. The ground gateways, serving as relay devices, form a homogeneous BPP. Due to the lower altitudes compared to satellites, they are approximately distributed on a sphere with radius $R_1 = 6371$km, which is the radius of the Earth. Relay satellites in multiple constellations distribute over $K-1$ concentric spheres of the Earth with fixed radii, $R_2,R_3 \dots R_K$, where $R_i<R_j$ for $i<j$. Since the satellites of different tiers belong to different constellations, we assumed that the satellites of each tier constitute a homogeneous BPP independently. Furthermore, the number and height of the devices in tier $k$ is denoted as $N_k$ and $h_k = R_k - R_1$, respectively. Hence, the height of gateways is assumed to be $h_1 = 0$. 

\par
Consider an ultra-distance communication between two far away points on the Earth requiring multi-hop relays. Each multi-hop routing starts with a ground transmitter and ends with another ground receiver. To further explain the restrictions in communication technique, we first give two definitions.
\begin{definition}[Dome Angle] 
The dome angle of two devices is the angle between two lines from two devices to the center of the Earth.
\end{definition}
\begin{definition}[Direction Angle]
Assume that relay devices A and B are on the same hop, and A transmits signal to the ground receiver through B. The direction angle is the angle between the line from B to A and the line from the receiver to A.
\end{definition}

\begin{figure}[h]
	\centering
	\includegraphics[width=0.98\linewidth]{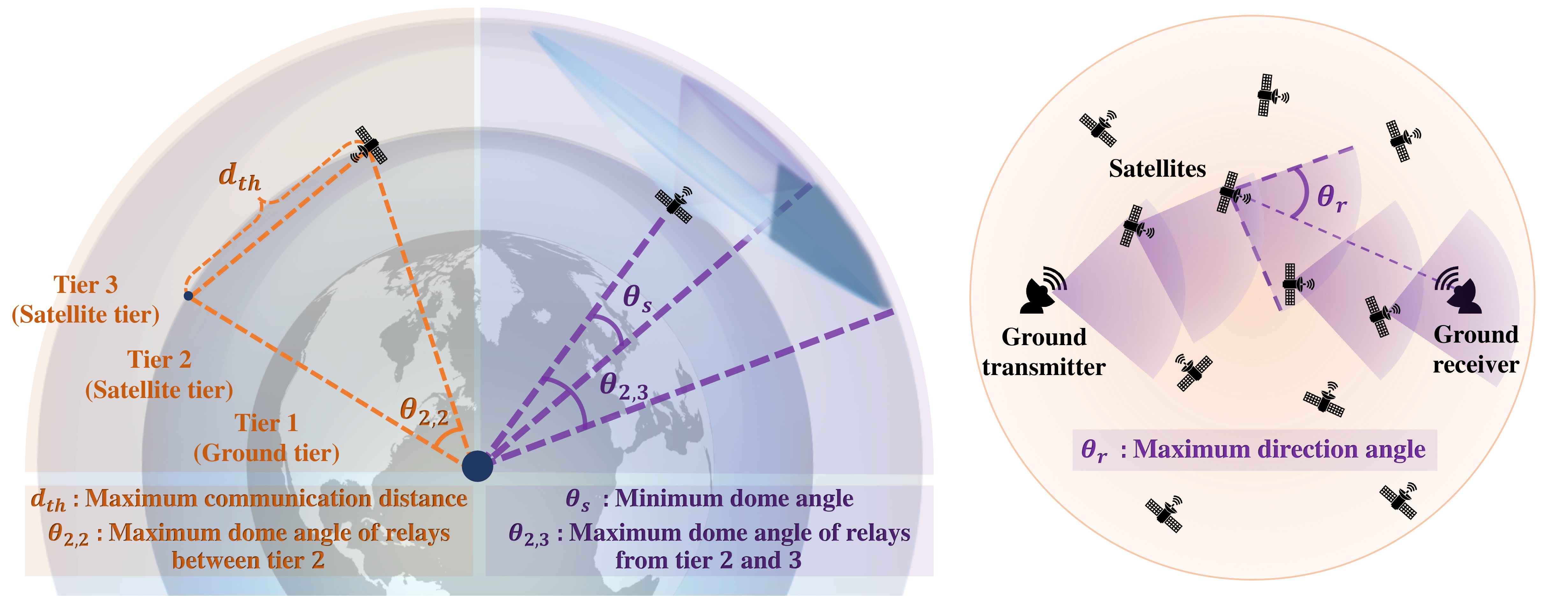}
	\caption{Schematic diagram of three constraints.}
	\label{fig:Figure2}
	\vspace{-0.4cm}
\end{figure}

As shown in Fig.~\ref{fig:Figure2}, the selected relay devices need to meet three constraints to ensure that the receiver is reachable: $\left(c_1\right)$ direction angle $\leq \theta_{r}$, $\left(c_2\right)$ dome angle $\geq \theta_{s}$, and  $\left(c_3\right)$ dome angle $\leq \theta_{i,j}$. Among devices that meet these restrictions, the device closest to the receiver will be selected as the relay for routing. The maximum direction angle $\theta_{r}$ is limited so that the next hop does not deviate significantly from the established direction. The right half of Fig.~\ref{fig:Figure2} is used for explaining $\theta_{r}$ from a two-dimensional projection perspective. The minimum dome angle $\theta_{s}$ ensures that the routing can reach its receiver within a limited number of hops, and the trajectory does not deviate significantly from the established path. The limitation of maximum dome angle comes from the attenuation of signal and blockage of the Earth. Assume that when the distance between two communication devices exceeds the threshold constant $d_{th}$, they cannot maintain reliable communication due to the attenuation. Furthermore, when two relay devices are in non-line-of-sight of each other, direct signals cannot propagate, and reliable communication cannot be guaranteed. Due to the blockage of the Earth, the dome angle between the two relays from tier $i$ and tier $j$ can not exceed $\arccos\left(\frac{R_1}{R_i}\right)+\arccos\left(\frac{R_1}{R_j}\right)$. Therefore, the maximum dome angle of relays from tier $i$ and $j$ is given by,
\begin{equation}\label{constraint 3}
    \theta_{i,j} = \max \left\{ \theta_{s} , \min \left\{ \arccos\left( \frac{R_i^2+R_j^2-d_{th}^2}{2 R_i R_j} \right) , \arccos\left(\frac{R_1}{R_i}\right)+\arccos\left(\frac{R_1}{R_j}\right) \right\} \right\}.
\end{equation}
It is necessary to ensure that the communication distance is smaller than $d_{th}$ and the transmission is not blocked by the Earth. However, when these two requirements are met, $\theta_{i,j}$ might be smaller than $\theta_{s}$, which means that communication devices from tier $i$ and $j$ cannot communicate directly within the same hop. In this case, $\theta_{s}$ can be set to the lower bound of $\theta_{i,j}$ for the sake of the subsequent derivation. In particular, when $i=j=1$, $2\arccos\left(\frac{R_1}{R_1}\right)=0$, and $\theta_{i,j} = \theta_{s}$, which is consistent with the fact that the two gateways cannot directly communicate over a long distance due to the blockage of the Earth. Finally, notice that the maximum direction angle $\theta_{r}$, the minimum dome angle $\theta_{s}$ and the maximum communication distance $d_{th}$ are pre-set. However, the maximum dome angle $\theta_{i,j}$ needs to be calculated by the height of two relay devices at the same hop.

\subsection{Single-hop Interruption Probability}
Based on the constraints from the previous section, the reliability of routing can be measured by interruption probability. 
\begin{definition}[Interruption Probability]
The interruption probability is the probability that no relay device is available in the region satisfying constraints $\left(c_1\right)$, $\left(c_2\right)$ and $\left(c_3\right)$.
\end{definition}
By definition, the interruption probability is negatively correlated with link reliability. In the following two lemmas, the tier to tier interruption probability and the total interruption probability of a single hop are given.

\begin{lemma}\label{lemma1}
The tier-to-tier interruption probability $P_{i,j}^{I}$ is the probability that a device in tier $i$ can not find any relays in tier $j$, which is given by
\begin{equation}\label{tier-to-tier}
P_{i,j}^{I} =
\left\{\begin{matrix}
    &\left(1- \frac{\theta_r}{4\pi}\left(\cos\theta_s - \cos\theta_{i,j}\right) \right)^{N_j} , &i \neq j \\
    &\left(1- \frac{\theta_r}{4\pi}\left(\cos\theta_s - \cos\theta_{i,i}\right) \right)^{N_i-1}, &i = j,
\end{matrix}\right.
\end{equation}
where $\theta_{i,j}$ is defined in (\ref{constraint 3}).
\begin{proof}
See appendix~\ref{app:lemma1}.
\end{proof}
\end{lemma}

All of the results for $1 \leq i,j \leq K$ in lemma~\ref{lemma1} can be recorded as matrix $P^{I}$. A more significant result is the probability of no available satellite within the given region of any tier. This probability is denoted as the single-hop interruption probability.

\begin{lemma}\label{lemma2}
For a device in tier $i$ needs to find a relay in any tier, the single-hop interruption probability $P_{i}^{S}$ is given by,
\begin{equation}\label{single-hop interruption}
    P_{i}^{S} = \left(1- \frac{\theta_r\left(\cos\theta_s - \cos\theta_{i,i}\right)}{4\pi} \right)^{N_i-1} \prod \limits_{j=1,j\neq i}^K \left(1- \frac{\theta_r\left(\cos\theta_s - \cos\theta_{i,j}\right)}{4\pi} \right)^{N_j},
\end{equation}
where $\theta_{i,j}$ is defined in (\ref{constraint 3}).
\begin{proof}
Since the distribution of relay devices between tiers is independent of each other, the probability that all tiers have no available relays is equal to the product of the probability that each tier has no available relays. Therefore, the result in (\ref{single-hop interruption}) is obtained.
\end{proof}
\end{lemma}

\section{Multi-hop Reliability Analysis}
\label{multi}
In this section, we extend the reliability result from single-hop to multi-hop. The connection between two adjacent hops is characterized by the transition probability matrix (TPM) in multi-hop routing. Based on the absorbing state and stationary distribution of the TPM and its variations, the multi-hop interruption probability is calculated, and the optimal priority strategy under stationary distribution is provided. Finally, we discuss the priority strategies of device selection at different tiers.

\subsection{TPM and Its Variations}\label{TPM}
When a single-hop network is changed to a multi-hop network, the relay devices in each tier have a different probability of being chosen. At the same time, the selection of the last hop directly affects the next hop. Therefore, the concepts of priority strategy and TPM are introduced. Consider that different tiers of relays have different priorities in selection. For example, when most relay satellites are located in one tier, devices in this tier can be used preferentially for routing to achieve a lower interruption probability. The following definition is given.

\begin{definition}[Priority Strategy]
$s$ is a vector whose $i_{th}$ element represents the priority of $i_{th}$ tier. The element's value is an integer ranging from 1 to $K$. A smaller value indicates a higher priority. The set of all priorities is denoted as $\mathcal{S}$.
\end{definition}

\par
For a given strategy $s$, a unique TPM $T^{(1)}$ is provided to calculate the probability of relays from different tiers being selected in each hop. For $1 \leq i,j \leq K$, element $T_{i,j}^{(1)}$ is the probability that the relay at tier $i$ selects device in tier $j$. $T_{i,j}^{(1)}$ is determined by the priority strategy $s$ and the tier-to-tier interruption probability $P_{i,j}^{I}$. The calculation process of $T^{(1)}$ through interruption probability and priority strategy is denoted as operator $\mathcal{T}^{(1)}$. The specific operation $T^{(1)} \leftarrow \mathcal{T}^{(1)}\left(s,P^{I}\right)$ is given as follows.

 \begin{algorithm}[!ht] 
	\caption{Algorithm of Operation $\mathcal{T}^{(1)}$.}
	\label{TPM1}
	\begin{algorithmic} [1]
		
		\STATE \textbf{Input}: Priority strategy vector $s$ and tier-to-tier interruption probability matrix $P^{I}$.
		
		\FOR{$i = 1 : K$}
		\FOR{$j = 1 : K$}
		
		\STATE $T_{i,j}^{(1)} \leftarrow \frac{1 - P_{i,j}^I}{1 - P_{i}^S}$.
		
		\FOR{$k = 1 : K$}

		\IF {$s_j > s_k$}
		\STATE $T_{i,j}^{(1)} \leftarrow T_{i,j}^{(1)} \cdot P_{i,k}^I$.
        \ENDIF
        
		\ENDFOR
		\ENDFOR
		\ENDFOR
		
		\STATE \textbf{Output}: TPM $T^{(1)}$.
	\end{algorithmic}
\end{algorithm}	

\par

Note that $P_i^{S}$ can be obtained from the elements in $P^{I}$, so it does not have to be an input. As a result, $T_{i,j}^{(1)}$ is the normalized probability that all tiers with a higher priority than tier $j$ have no available relays, and there are available relays at tier $j$. Normalization makes the sum of each row of the matrix equal to 1, which is the characteristic of TPM. However, the resulting $T^{(1)}$ does not explicitly provide the interruption probability. Therefore, we introduce the absorbing state and add a column and a row to store the absorbing state. The definition of absorbing state and the specific operation $\widetilde{T}^{(2)} \leftarrow \widetilde{\mathcal{T}}^{(2)}\left(s,P^{I}\right)$ are given as follows, where $\widetilde{T}^{(2)}$ is the augmented TPM.
\begin{algorithm}[!ht] 
	\caption{Algorithm of Operation $\widetilde{\mathcal{T}}^{(2)}$.}
	\label{TPM2}
	\begin{algorithmic} [1]
		
		\STATE \textbf{Input}: Priority strategy vector $s$ and tier-to-tier interruption probability matrix $P^{I}$.
		
		\FOR{$i = 1 : K$}
		\FOR{$j = 1 : K$}
		
		\STATE $\widetilde{T}_{i,j}^{(2)} \leftarrow 1 - P_{i,j}^I$.
		\STATE Execute steps (5)-(9)	in algorithm~\ref{TPM1} (substitute $T^{(1)}_{i,j}$ into $\widetilde{T}_{i,j}^{(2)}$).
        
		\ENDFOR
		
		\STATE $\widetilde{T}_{i,K+1}^{(2)} \leftarrow P_{i}^{S}$,  $\widetilde{T}_{K+1,i}^{(2)} \leftarrow 0$.
		\ENDFOR
		\STATE $\widetilde{T}_{K+1,K+1}^{(2)} \leftarrow 1$.
		
		\STATE \textbf{Output}: Augmented TPM $\widetilde{T}^{(2)}$.
	\end{algorithmic}
\end{algorithm}

\begin{definition}[Absorbing State]
The absorbing state of TPM represents the state where the routing is interrupted because of none available relays.
\end{definition}

\begin{figure}[h]
	\centering
	\includegraphics[width=0.75\linewidth]{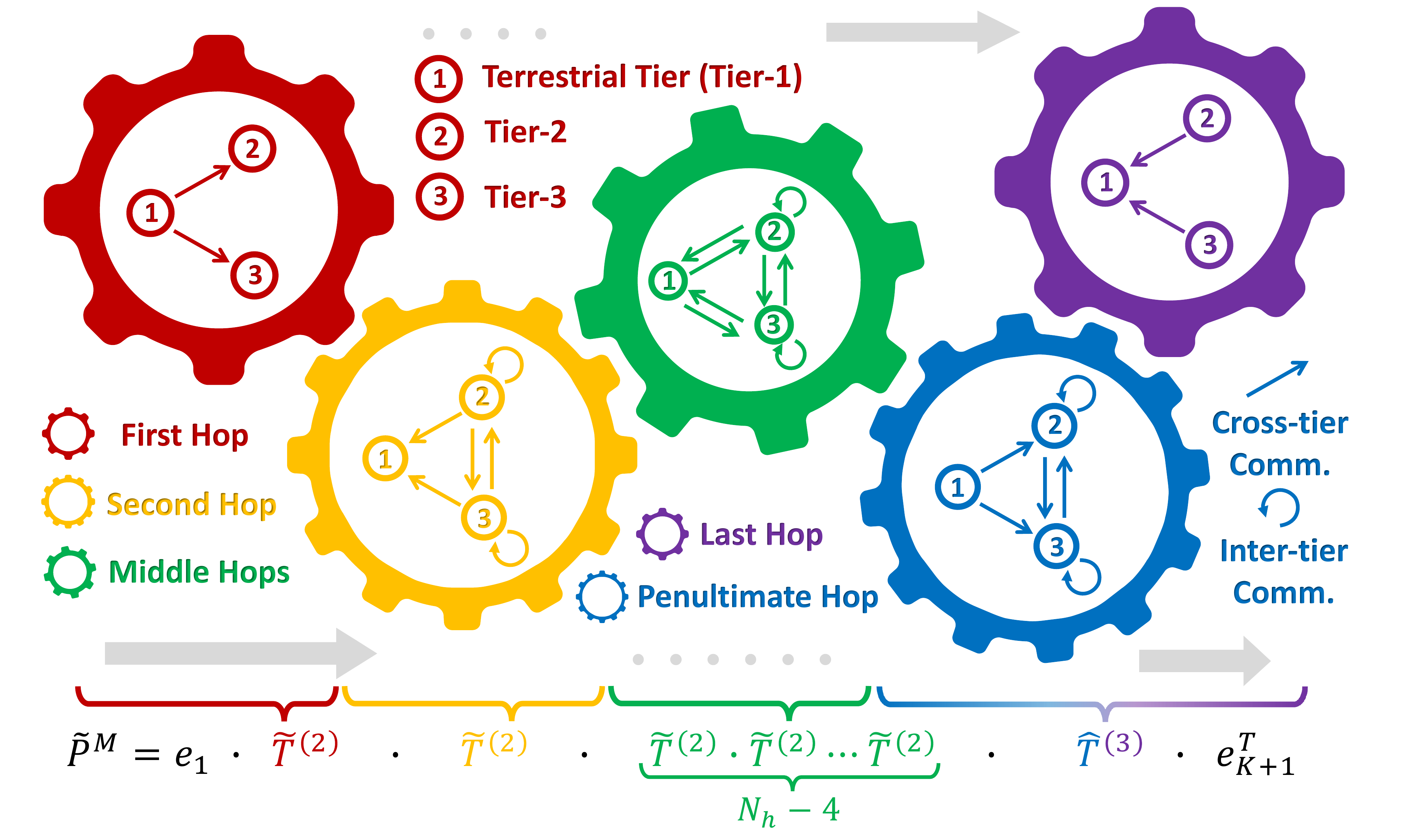}
	\caption{Example of the tier choice for a three-tier network.}
	\label{fig:Figure3}
	\vspace{-0.4cm}
\end{figure}

Using the augmented matrix $\widetilde{T}^{(2)}$, the interruption probability of each hop can be obtained by matrix operation. However, both of the matrices $T^{(1)}$ and $\widetilde{T}^{(2)}$ apply to the hops other than the last two hops. As shown in Fig.~\ref{fig:Figure3}, assume the third tier communicates directly with the first (terrestrial) tier. The state transition diagram is bidirectionally connected only in the middle hops. Since the ground relay cannot communicate with the ground receiver, if the relay of the penultimate hop is selected at the first tier, the communication will interrupt. Therefore, we provide another variation of augmented matrix $\widehat{T}^{(3)}$. The specific operation $\widehat{T}^{(3)} \leftarrow \widehat{\mathcal{T}}^{(3)}\left(s,P^{I}\right)$ is as follows.
\par

\begin{algorithm}[!ht] 
	\caption{Algorithm of Operation $\widehat{\mathcal{T}}^{(3)}$.}
	\label{TPM3}
	\begin{algorithmic} [1]
		
		\STATE \textbf{Input}: Priority strategy vector $s$ and tier-to-tier interruption probability matrix $P^{I}$.
		
		\FOR{$i = 1 : K$}
		\FOR{$j = 1 : K$}
		
        \STATE $\widehat{T}_{i,j}^{(3)} \leftarrow \left(1 - P_{i,j}^I\right) \mathbbm{1} \{ P_{j,1}^I \neq 1 \}$.
        
		\FOR{$k = 1 : K$}
		\IF {$s_j > s_k$ and $P_{k,1}^I \neq 1$}
		\STATE $\widehat{T}_{i,j}^{(3)} \leftarrow \widehat{T}_{i,j}^{(3)} \cdot P_{i,k}^I$.
        \ENDIF
        
		\ENDFOR
		\ENDFOR
		
		\STATE $\widehat{T}_{i,K+1}^{(3)} \leftarrow 1 - \sum_{k=1}^{K} \widehat{T}_{i,k}^{(3)}$, \ $\widehat{T}_{K+1,i}^{(3)} \leftarrow 0$.
		\ENDFOR
		\STATE $\widehat{T}_{K+1,K+1}^{(3)} \leftarrow 1$.
		
		\STATE \textbf{Output}: Augmented TPM $\widehat{T}^{(3)}$.
	\end{algorithmic}
\end{algorithm}

In step (4) of algorithm~\ref{TPM3}, $\mathbbm{1} \{ \cdot \}$ is an indicator function. When $P_{j,1}^I \neq 1$, $\mathbbm{1} \{ P_{j,1}^I \neq 1 \}=1$, otherwise $\mathbbm{1} \{ P_{j,1}^I \neq 1 \}=0$. Compared to operation $\widetilde{T}^{(2)} \leftarrow \widetilde{\mathcal{T}}^{(2)}\left(s,P^{I}\right)$, operation $\widehat{T}^{(3)} \leftarrow \widehat{\mathcal{T}}^{(3)}\left(s,P^{I}\right)$ moves the priority of tiers that cannot communicate with the first tier to last, and set the transition probability toward those tiers as 0, since selecting the last relay in tiers that cannot communicate directly with the first tier is unreasonable. Furthermore, although state transition diagrams are not bidirectional in the first two hops, $\widetilde{T}^{(2)}$ is also applicable.

\subsection{Multi-hop Interruption Probability}
In this subsection, several important analytical conclusions are derived: the average number of hops before interruption, the average number of hops required for successful transmission, and the multi-hop interruption probability. The average number of hops before interruption, which can also be regarded as the average number of hops before the absorbing state occurs, can be calculated by the following lemma.

\begin{lemma}\label{mu_1}
Given that the route starts at the $i^{th}$ tier, the average number of hops before interruption, denoted as $\mu_i$, is derived by solving the following equations,
\begin{equation} \label{mu}
    \mu_i = 1 + \sum_{j \in \mathcal{K}} \widetilde{T}_{i,j}^{(2)} \mu_j,\ i \in \mathcal{K},
\end{equation}
where $\mathcal{K}$ is the index set of reachable tiers starting from the $i^{th}$ tier, the augmented matrix $\widetilde{T}^{(2)}$ is derived by $\widetilde{T}^{(2)} \leftarrow \widetilde{\mathcal{T}}^{(2)}\left(s,P^{I}\right)$ and interruption probability matrix $P^I$ is defined in (\ref{tier-to-tier}). 
\begin{proof}
See appendix~\ref{app:mu_1}.
\end{proof}
\end{lemma}

Note that reachable tiers in $\mathcal{K}$ also include indirectly reachable tiers. For example, if a ground gateway can communicate with other ground gateways through a relay satellite, the first tier is a reachable tier starting from the first tier, even if the gateways cannot communicate directly with each other. Then, we want to estimate the average number of hops required for successful transmission, denoted as $N_h$. The biggest challenge of deriving $N_h$  comes from changing the probability of relay device selection at each tier during routing. Thus, the stationary distribution is introduced.

\begin{definition}[Stationary Distribution]
The stationary distribution $v$ is a steady-state of the probability of relay device selection at each tier. It can be expressed as the left eigenvector of TPM whose eigenvalue is 1, that is, $v T^{(1)} = v$.
\end{definition}

The stationary distribution will provide weighting for the calculation of $N_h$. The following lemma can provide an accurate estimate for a low interruption probability.

\begin{lemma}\label{lemma3}
In the direction of the shortest path routing, the approximate average number of hops required for successful transmission is derived by,
\begin{equation}\label{N_h}
    N_h = \left[ \frac{\theta_m}{\overline{\theta}_o} \right],
\end{equation}
where $[\cdot]$ represents round to an integer, $\theta_{m}$ is the dome angle between the transmitter and receiver, the average dome angle $\overline{\theta}_o$ is given by,
\begin{equation}\label{theta_average}
\begin{split}
    \overline{\theta}_o &= \sum_{i \in \mathcal{K}} v_i \sum_{j \in \mathcal{K},j \neq i} T_{i,j}^{(1)} \arccos\left( \frac{2 \pi}{\theta_r} - \frac{2 \pi}{\theta_r} \cos\left(\pi \prod_{k=1}^{N_j}\frac{2k-1}{2k} \right) + \cos\theta_{i,j} \right)\\
    & + \sum_{i \in \mathcal{K}} v_i T_{i,i}^{(1)} \arccos\left(  \frac{2 \pi}{\theta_r} - \frac{2 \pi}{\theta_r} \cos\left(\pi \prod_{k=1}^{N_i-1}\frac{2k-1}{2k} \right) + \cos\theta_{i,i} \right),
\end{split}
\end{equation}
where $\theta_{i,j}$ is defined in (\ref{constraint 3}), $T_{i,j}^{(1)}$ is derived by $T^{(1)} \leftarrow \mathcal{T}^{(1)}\left(s,P^{I}\right)$, and $v$  is the stationary distribution, which is obtained by $v \, T^{(1)}=v$, $v_i$ is the $i^{th}$ element of $v$, $\mathcal{K}$ is the index set of reachable tiers.
\begin{proof}
See appendix~\ref{app:lemma3}.
\end{proof}
\end{lemma}

An obvious corollary is that multi-hop routings are reliable when $\mu_1$ is much larger than $N_h$. Based on TPM and its variations and $N_h$, the multi-hop interruption probability is given as follows. Note that $N_h$ is approximate, so the multi-hop interruption probability is also an approximation.

\begin{theorem}\label{theorem1}
The approximate multi-hop interruption probability $\widetilde{P}^M$ in the direction of the shortest path routing is derived by,
\begin{equation}\label{p_I}
    \widetilde{P}^M = e_1 \left(\widetilde{T}^{(2)}\right)^{N_h-2} \widehat{T}^{(3)} e_{K+1}^T,
\end{equation}
where $e_i$ is standard unit row vector with $K+1$ elements where only the $i^{th}$ element is 1 and all other $K$ elements are 0, $e_{K+1}^T$ is the transpose of row vector $e_{K+1}$. The augmented TPMs in (\ref{p_I}) are obtained by $\widetilde{T}^{(2)} \leftarrow \widetilde{\mathcal{T}}^{(2)}\left(s,P^{I}\right)$ and $\widehat{T}^{(3)} \leftarrow \widehat{\mathcal{T}}^{(3)}\left(s,P^{I}\right)$ and interruption probability matrix $P^I$ is defined in (\ref{tier-to-tier}).
\begin{proof}
See appendix~\ref{app:theorem1}.
\end{proof}
\end{theorem}

The following proposition is given as a supplement to the theorem. It shows that even if $\widetilde{T}^{(2)}$ and $\widehat{T}^{(3)}$ are not unique, the multi-hop interruption probability $\widetilde{P}^M$ is still unique.

\begin{proposition}\label{proposition2}
The approximate multi-hop interruption probability $\widetilde{P}^M$ in (\ref{p_I}) is unique.
\begin{proof}
See appendix~\ref{app:proposition2}.
\end{proof}
\end{proposition}

\begin{remark}[Remarks of Theorem~\ref{theorem1}]
The following remarks are worth mentioning based on Theorem~\ref{theorem1}.
\begin{itemize}
    \item The transmitter and receiver do not have to locate in the first tier. Only minor adjustments are required to extend the transmitter and receiver in any tier.
    \item In general, the probabilities of locating on different tiers will soon converge to the stationary distribution. In the long term, the tiers where the transmitter and receiver are located have little effect on the results of Theorem~\ref{theorem1}.
    \item Other adaptive or intelligent strategies, such as dynamic directional angle strategy, can be applied to further prevent the occurrence of communication interruption. Therefore, even the interruption probability derived by Theorem~\ref{theorem1} with the best priority strategy is not equivalent to the minimum interruption probability in the actual situation.
\end{itemize}
\end{remark}

According to Theorem~\ref{theorem1}, interruption probability is related to priority strategy. Therefore, the optimal strategy based on stationary distribution is studied in the next subsection.

\subsection{Stationary Optimal Strategy}
This subsection discusses the potential optimal priority strategy. To start with, we need to set a guideline for evaluating whether a priority strategy is reliable or not. An intuitive idea is evaluated by interruption probability. However, for a given strategy, the probability that the relay device is located at each tier in different hops is different, which results in a change in the interruption probability of each hop. Therefore, the following proposition is provided to handle this problem.

\begin{proposition}\label{proposition1}
When all tiers of the network are reachable, for a given priority strategy, we can get a unique stationary distribution $v$. For a given stationary distribution, we can get a unique weighted single-hop interruption probability $\overline{P}^S$, which is defined as,
\begin{equation}\label{weighted single-hop}
    \overline{P}^S = \sum_{i=1}^K v_i P_i^S,
\end{equation}
where $v$ is the stationary distribution obtained by $v \, T^{(1)}=v$, $v_i$ is the $i^{th}$ element of $v$ and $P_{i}^S$ is the single-hop interruption probability, which is defined in (\ref{single-hop interruption}).
\begin{proof}
See appendix~\ref{app:proposition1}.
\end{proof}
\end{proposition}

Notice that even if there are some unreachable tiers, we can remove the rows and columns corresponding to unreachable tiers and get a submatrix, whose stationary distribution is still unique. So far, we have built a bridge between strategies and the interruption probability. Then, a potential optimal priority strategy is provided below.

\begin{definition}[Stationary Optimal Priority Strategy]
The stationary optimal priority strategy $s^* \in \mathcal{S}$ is the priority strategy with the weighted single-hop interruption probability $\overline{P}^S$ defined in (\ref{weighted single-hop}).
\end{definition}

The algorithm of searching for the stationary optimal priority strategy $s^*$ is given. By definition, the algorithm needs to traverse all the strategies $s \in \mathcal{S}$. After each strategy selection, the stationary distribution of $T^{(1)}$ is obtained, and the single-hop interruption probability is obtained by stationary distribution and $\widetilde{T}^{(2)}$. Through the above analysis, the algorithm is provided.

\begin{algorithm}[!ht] 
	\caption{Stationary Optimal Priority Strategy Searching Algorithm.}
	\begin{algorithmic} [1]
		\label{alg:s*}
		\STATE \textbf{Input}: Tier-to-tier interruption probability matrix $P^{I}$.
		
        \STATE $p_{\min} \leftarrow 1$.
		\REPEAT
		\STATE Select an $\widehat{s} \in \mathcal{S}$ which has not been chosen.
		
		\STATE $T^{(1)} \leftarrow \mathcal{T}^{(1)} \left(\widehat{s},P^I\right)$.
		\STATE Obtain $v_{K \times 1}$ by equation $v T^{(1)} = v$.
		\STATE $\widetilde{v}_i \leftarrow v_i$ for $\forall i \leq K$ and $\widetilde{v}_{K+1} \leftarrow 0$.
		\STATE $\widetilde{T}^{(2)} \leftarrow \widetilde{T}^{(2)} \left(\widehat{s},P^I\right)$.
		\STATE $w \leftarrow \widetilde{v} \widetilde{T}^{(2)}$.
		
		\IF {$w_{K+1} < p_{\min}$}
		\STATE $p_{\min} \leftarrow p_{K+1}$.
		\STATE $s^* \leftarrow \widehat{s}$.
        \ENDIF
		
		\UNTIL{All of the strategies $\widehat{s} \in \mathcal{S}$ are selected.}
		
		\STATE \textbf{Output}: Stationary optimal priority strategy $s^*$.
	\end{algorithmic}
\end{algorithm}	


\section{Case Study of A Three-Tier HSTRN}
In this section, we consider a scenario in which Kepler and OneWeb constellations complete multi-hop routing in cooperation with the terrestrial network. The terrestrial tier is composed of 300 gateways. The numbers of satellites with heights of $575~$km and $1200~$km are 140 and 720, as imitations of Kepler and OneWeb constellations, respectively \cite{petit2021assessment}.

\subsection{Simulation Procedure}\label{simulation}
Monte Carlo simulation is one of the most common methods to verify the accuracy of analytical results under stochastic geometry framework. In our simulation, all of the priority strategies $s \in \mathcal{S}$ are traversed to find $s^*$. For each strategy, we randomly generate the positions of devices and perform multi-hop routing for $10^6$ iterations. The ground transmitter and receiver are on opposite sides of the Earth, that is, $\theta_{m} = \pi$. As stated earlier, the closest relay to the receiver is preferred each time the next hop is selected within the constraint region dynamically hop by hop.

\begin{figure}[h]
	\centering
    \includegraphics[width=0.8\linewidth]{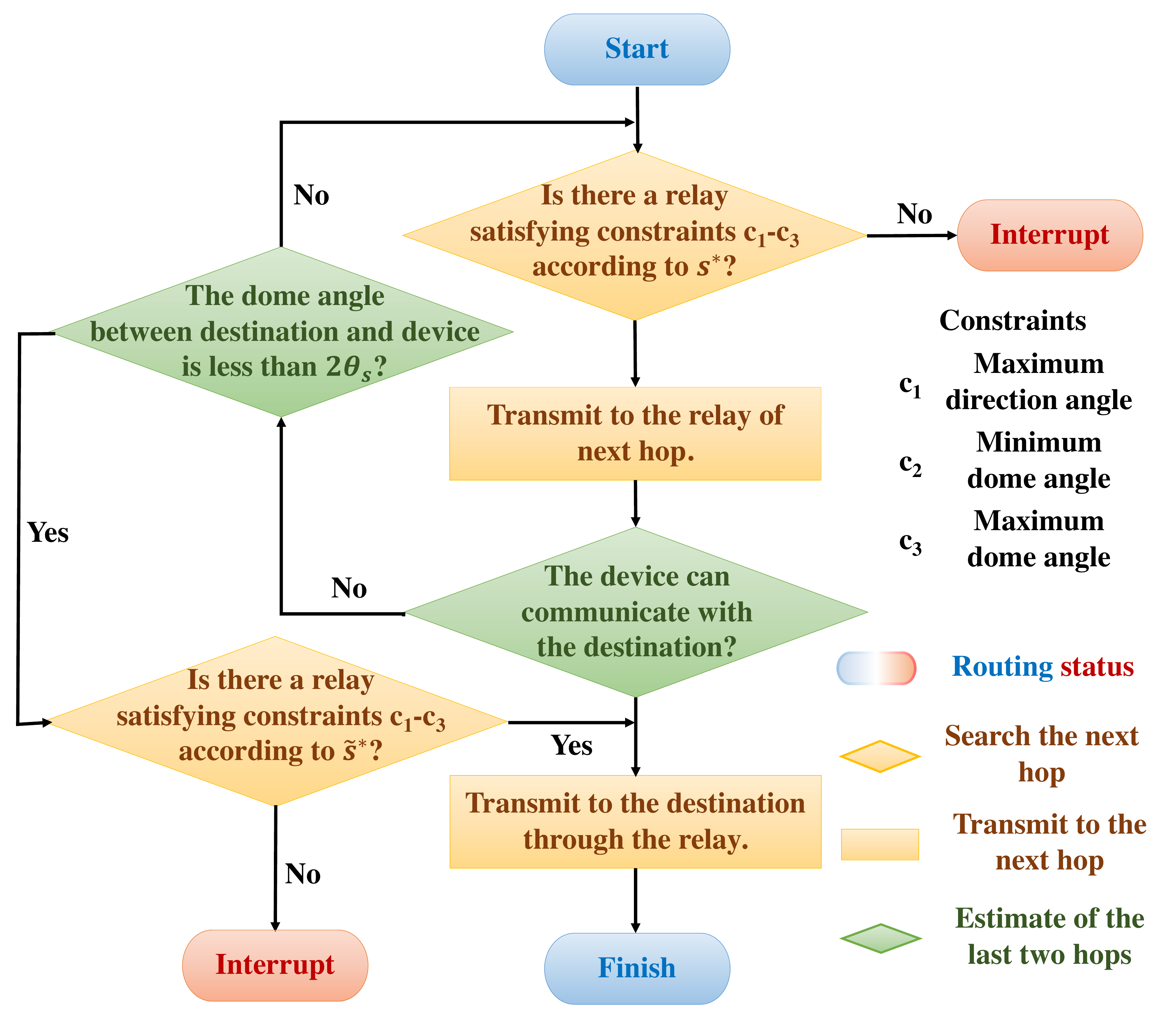}
	\caption{Multi-hop routing simulation flow chart.}
	\label{fig:Figure4}
	\vspace{-0.4cm}
\end{figure}

The parameter of constraints are the maximum direction angle $\theta_r = \frac{\pi}{6}$, the minimum dome angle $\theta_s = \frac{\pi}{10}$, and the maximum reliable communication distance $d_{th} = 4000$ km. The simulation flow chart of routing is shown in Fig.~\ref{fig:Figure4}. To ensure that the devices of the last hop can communicate with the receiver, $s^*$ in the penultimate hop is adjusted slightly as $\widetilde{s}^*$, whose priority of the first tier is moved to the end. 

\subsection{Discussion of Priority Strategies}\label{discussion of priority}
In this section, we calculate the notations' values according to the parameters given in the case study. Based on these numerical results, the rationality of the stationary optimal priority strategy is proved, and other potentially feasible strategies are discussed.

\par
Intuitively, the third tier has the highest density of satellites per unit area, while the second tier has the lowest density (i.e., $s=[3~1~2]$). We can call this naive priority strategy the density inspired strategy. However, this strategy usually fails to achieve the minimum interruption probability, and this case study is an example. The tier-to-tier interruption probability matrix $P^I$ defined in \eqref{tier-to-tier}:
\begin{equation}
 P^I =  \begin{bmatrix}
1.0000    & 0.8208 & 0.0466 \\ 
0.6549&  0.5074& 0.0503\\ 
 0.2787  &  0.5591& 0.0659
\end{bmatrix},
\end{equation}
where $P_{i,j}^{I}$ is the probability that a device in tier $i$ can not find any relays in tier $j$. Then, the single-hop interruption probability $P^S$ defined in \eqref{single-hop interruption} is 
\begin{equation}
    P^S = \begin{bmatrix} 0.0383 & 0.0166 & 0.0102 \end{bmatrix}.
\end{equation}
As shown in $P^S$, it is much harder to find the next hop at the ground tier than at the second tier. Compared to the density inspired strategy, a smarter priority strategy is $s=[3~2~1]$. The strategy determined by $P^S$ is called the single-hop interruption probability inspired strategy.

\par
Then, we compare the TPM $T^{(1)}$ and its variations $\widetilde{T}^{(2)}$, $\widehat{T}^{(3)}$ through their values of elements: 
\begin{equation}
\begin{split}
T^{(1)} = \begin{bmatrix}
 0 &  0.0087 &  0.9913\\ 
 0.0089 &  0.0253  & 0.9658\\ 
 0.0267 &  0.0292& 0.9440
\end{bmatrix}, \ \ \ \ \  \ \ \ \ \ \ \ \ \ \ \ \ \ \ \ \ \ \ \ \  \\
\widetilde{T}^{(2)}= \begin{bmatrix}
         0 &   0.0084  &  0.9534  &  0.0383 \\
    0.0088  &  0.0249  &  0.9497  &  0.0166 \\
    0.0265  &  0.0289  &  0.9344  &  0.0102 \\
         0   &      0  &       0  & 1.0000
\end{bmatrix}, \
 \widehat{T}^{(3)} =  \begin{bmatrix}
         0  &  0.0084 &   0.9534  &  0.0383 \\ 
         0   & 0.0249 &   0.9497  &  0.0254 \\
         0   & 0.0289 &   0.9344  &  0.0367 \\
         0    &     0 &        0  &  1.0000
\end{bmatrix}.
\end{split}
\end{equation}
It can be shown that $T^{(1)}$ is obtained from the row normalization of the first three rows and columns of $\widetilde{T}^{(2)}$. The first three elements in the last column of $\widetilde{T}^{(2)}$ are $P^S$ transpose. Comparing $\widetilde{T}^{(2)}$ and $\widehat{T}^{(3)}$, since the first tier has the lowest priority, the second and third columns of the two matrices are the same. The values of the fourth column in $\widehat{T}^{(3)}$ are calculated by the sum of the values of the first column and the fourth column in $\widetilde{T}^{(2)}$.

\par
Then, we introduce the stationary optimal priority strategy based on the stationary distribution. Compared to the previous two strategies, it is more carefully designed in algorithm~\ref{alg:s*} and therefore expected to have better reliability performance. The augmented stationary distribution $\widetilde{v}$ and interruption probability $w$ under different strategies are given in table~\ref{sim_1}. Under the stationary distribution, the probabilities of staying at three tiers after one hop are the values of $w$'s first three elements. The last element of $w$ is the single-hop interruption probability under the stationary distribution. The rightmost column in table~\ref{sim_1} is the real interruption probability $P^M$ obtained by simulation. Simulation shows that $s = [3 ~2~ 1]$ is the optimal strategy and the stationary optimal priority strategy gets the right answer. A more meaningful consequence is that the smaller the actual interruption probability $P^M$, the smaller the last element of $w$. In other words, the ranking of strategies provided by simulation and the stationary optimal priority strategy is exactly the same. Last but not least, a further comparison of the strategies is presented in Sec.~\ref{alternatives}.

\begin{table}[t]
\caption{The interruption probabilities under different strategies.}
\label{sim_1}
\centering
\renewcommand\arraystretch{1.2}
\begin{tabular}{|c|c|c|c|}
\hline
Strategy & Stationary Distribution    $\widetilde{v}$      &  $w = \widetilde{v} \widetilde{T}^{(2)}$    & $P^M$ (Simu.)   \\  \hline  \hline
$[3   ~    2   ~  1]$   & $[0.0255   ~   0.0286  ~    0.9459    ~     0]$ & $[0.0253  ~    0.0283  ~  0.9353    ~  0.0111]$ & 0.1033 \\ \hline
$[2   ~  3   ~    1]$   & $[0.0454  ~    0.0082 ~   0.9464   ~        0]$ & $[0.0449  ~    0.0081 ~   0.9354  ~    0.0116]$ & 0.1122 \\ \hline
$[3   ~    1    ~ 2]$   & $[0.0179  ~    0.4680  ~  0.5141     ~      0]$ & $[0.0177   ~   0.4616 ~   0.5070    ~  0.0137]$ & 0.1155 \\ \hline
$[2   ~    1   ~  3]$   & $[0.2221   ~   0.4118  ~    0.3661   ~      0]$ & $[0.2194   ~   0.4051  ~    0.3564  ~  0.0191]$ & 0.2135 \\ \hline
$[1   ~    3   ~  2]$   & $[0.4197   ~   0.0084  ~    0.5719   ~      0]$ & $[0.4154  ~    0.0083  ~  0.5543     ~ 0.0220]$ & 0.3417 \\ \hline
$[1  ~   2   ~   3]$    & $[0.3809   ~   0.1861  ~    0.4330   ~      0]$ & $[0.3766  ~    0.1818   ~ 0.4195  ~    0.0221]$ & 0.3432 \\ \hline
\end{tabular}
\end{table}


\subsection{Verification of Analytical Results}
\begin{figure*}[htbp]
\begin{minipage}[t]{0.48\linewidth}
\centering
\includegraphics[width=\linewidth]{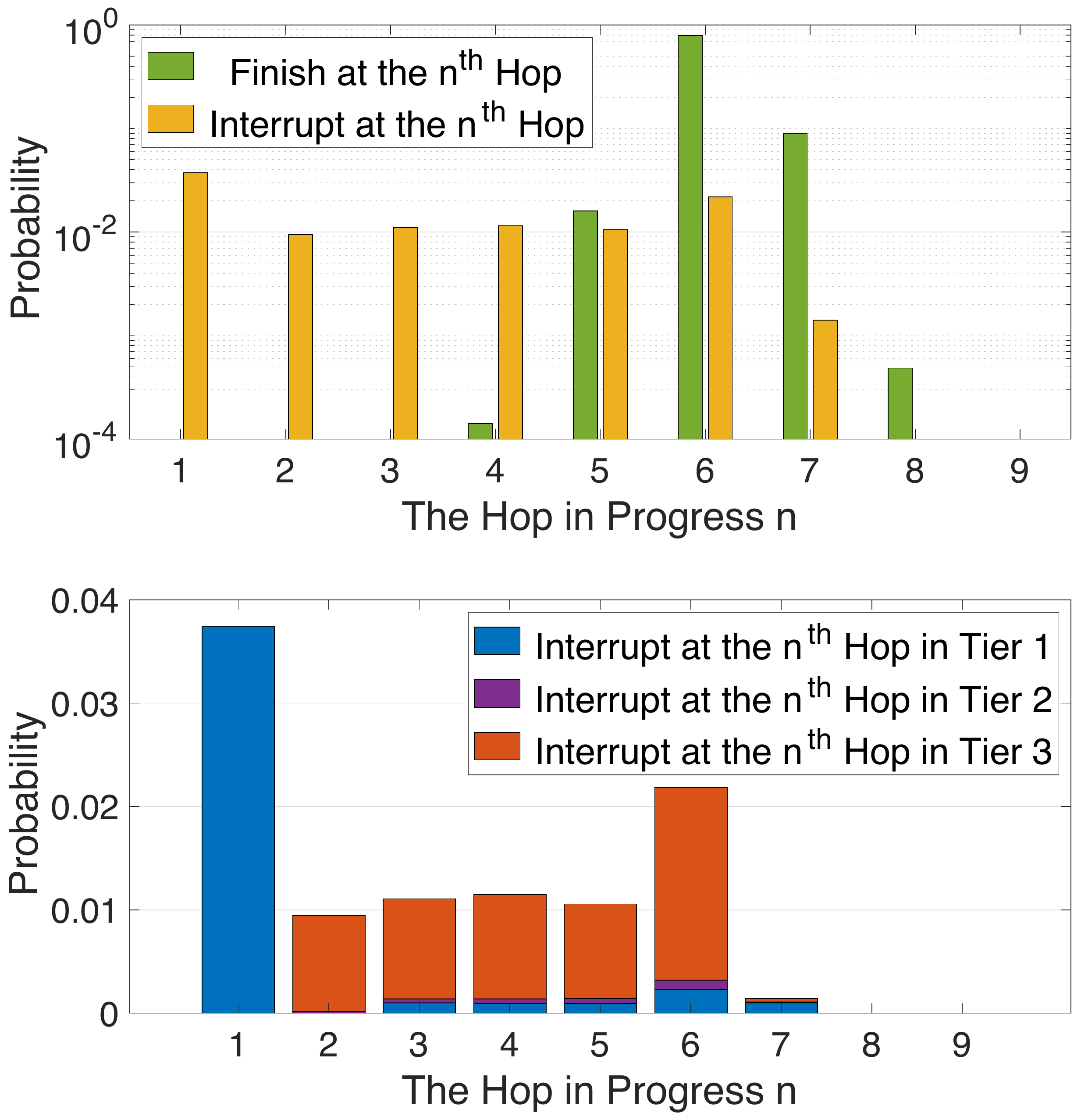}
\caption{Number of hops before the interruption and successful transmission.}
\label{fig:Figure10}
\end{minipage}
\hfill
\begin{minipage}[t]{0.48\linewidth}
\centering
\includegraphics[width=\linewidth]{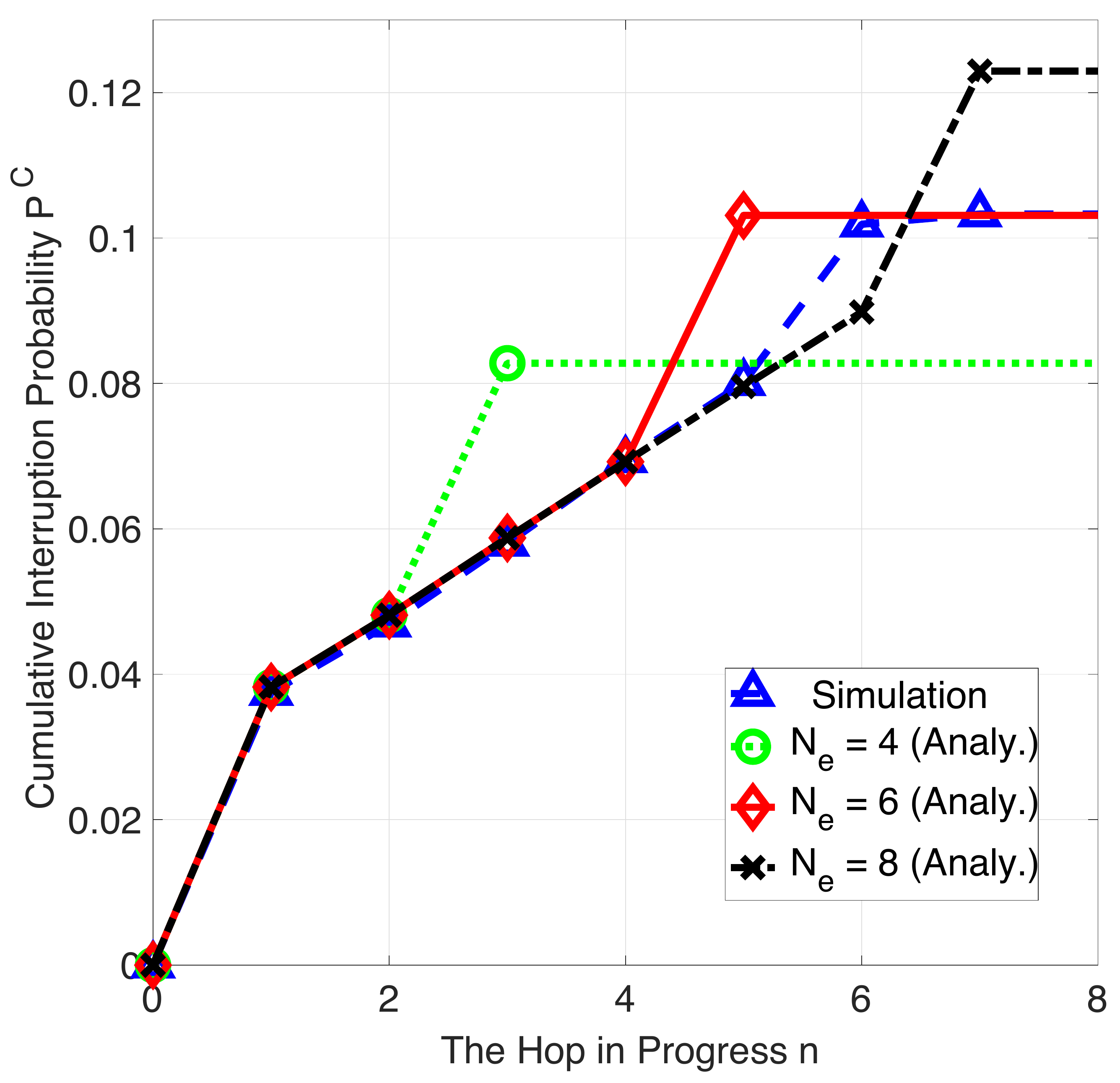}
\caption{Graph of cumulative interruption probability with different hops in progress.}
\label{fig:Figure5}
\end{minipage}
\end{figure*}

In this subsection, the accuracy of analytical results in section \ref{Single} and section \ref{multi} is verified. As shown in the top half of Fig.~\ref{fig:Figure10}, most routes end in the sixth hop (notice that the vertical axis is logarithmic), and the average number of hops required for successful transmission is 6.08. In our analytical framework, the average dome angle in \eqref{theta_average} is $\overline{\theta}_o = 0.4915$, and the average number of hops required for successful transmission in \eqref{N_h} is $N_h = 6$, which is highly consistent with the simulation result. As defined in \eqref{mu}, the average number of hops before an interruption is $\mu = [87.516~89.4314~89.9615]$, which shows that $\mu$ is less affected by the starting tier. Since the values of elements in $\mu$ are much larger than the estimated $N_h = 6$, the route most likely will not be interrupted.

\par
Next, we verify the accuracy of the multi-hop interruption probability given in Theorem~\ref{theorem1}. The cumulative interruption probability $P^C \left(n,N_e\right)$ 
 in Fig.~\ref{fig:Figure5} is given as,
\begin{equation}
    P^C \left(n,N_e\right) = 
\left\{\begin{matrix}
    &e_1 \left(\widetilde{T}^{(2)}\right)^{n} e_{K+1}^T , &n < N_e-1, \\
    &e_1 \left(\widetilde{T}^{(2)}\right)^{N_e-2} \widehat{T}^{(3)} e_{K+1}^T, &n \geq N_e-1,
\end{matrix}\right.
\end{equation}
where $N_e$ is the predicted number of hops required to finish the routing, and it is also shown in the label of Fig.~\ref{fig:Figure5}, $n$ is the hop in progress, and the definitions of remaining parameters are given in Theorem~\ref{theorem1}.  $P^C \left(n,N_e\right)$ can be regarded as the estimation of the probability of an interruption occurring before $n$ hops. Fig.~\ref{fig:Figure5} shows the analysis results with $N_e=4,6,8$. When $N_e=N_h=6$, $\widetilde{P}^M=P^C \left(n,6\right)$ stopped at 0.1031, and the simulation result also stopped at 0.1033. Therefore, Fig.~\ref{fig:Figure5} proves that multi-hop interruption probability estimated by Theorem~\ref{theorem1} is accurate and in accordance with the simulation. In fact, not only the interruption probability of the entire route, the interruption probability of each hop is possible to be predicted by the analytical expression in Theorem~\ref{theorem1}. Furthermore, whether from the bottom half of Fig.~\ref{fig:Figure10} or the slope of lines in Fig.~\ref{fig:Figure5}, we can find that the interrupt probabilities of the middle hops do not change much, while the first and last hops are more likely to be interrupted.

\section{Further Extension}
In this section, We further discuss the multi-hop reliability performance under different strategies and network deployments. Unless otherwise stated, the parameters of constraints are $\theta_r = \frac{\pi}{6}$, $\theta_s = \frac{\pi}{10}$, $d_{th} = 4000$ km, and the dome angle between the transmitter and the receiver is $\theta_m = \pi$. In addition, by adjusting the constraints, the existing reliability analysis framework can be extended to the analysis of availability, coverage probability, and URLLC rate. Finally, more intelligent routing schemes and variants of point processes in HSTRN modeling are discussed.

\subsection{Alternatives for Stationary Optimal Strategy}\label{alternatives}
We start this subsection with the motivation for finding alternatives. For the first three algorithms, as the number of tiers increases, the computation time of the algorithm increases at a cubic speed, which means that the algorithm complexity is $\mathcal{O}(K^3)$. Generally speaking, the number of tiers is small, and this complexity is acceptable. However, for each priority strategy, algorithm~\ref{alg:s*} needs cubic computation complexity. According to the whole permutation formula of the ordered sequence, there are $K!$ priority strategies for $K$ tiers. An unfortunate result is that the complexity of algorithm~\ref{alg:s*} is $\mathcal{O}(K^3 \cdot K!)$. Therefore, we hope to find several alternatives to this computationally expensive strategy. Recall that in Sec.~\ref{discussion of priority}, two potential priority strategies are mentioned. Next, we will list several priority strategies and compare their reliability performance.

\begin{itemize}
    \item Exhaustive search: Through the simulation method in Sec.~\ref{simulation}, all of the strategies are traversed by exhaustive search, and the one with the lowest interruption probability is selected. Although this strategy provides the lowest interruption probability and is theoretically optimal, it is often difficult to do $10^6$ rounds of pre-testing for each strategy in practice.
    \item Stationary optimal priority strategy: The strategy $s^*$ obtained by algorithm~\ref{alg:s*}.
    \item Single-hop interruption probability inspired strategy: An alternative of $s^*$. Through single-hop interruption probability defined in (\ref{single-hop interruption}), give higher priority to tiers with lower single-hop interruption probability. 
    \item Density inspired strategy: An alternative of $s^*$. Give the tiers with indices $i = 1$ or $i \not\in \mathcal{K}$ the lowest priority. For the tiers with indices $i \in \mathcal{K}$, the higher the density of satellites, the higher the priority. The density is calculated by $\frac{N_i}{4 \pi R_i^2}$.
\end{itemize}

\begin{figure*}[htbp]
\begin{minipage}[t]{0.48\linewidth}
\centering
\includegraphics[width=\linewidth]{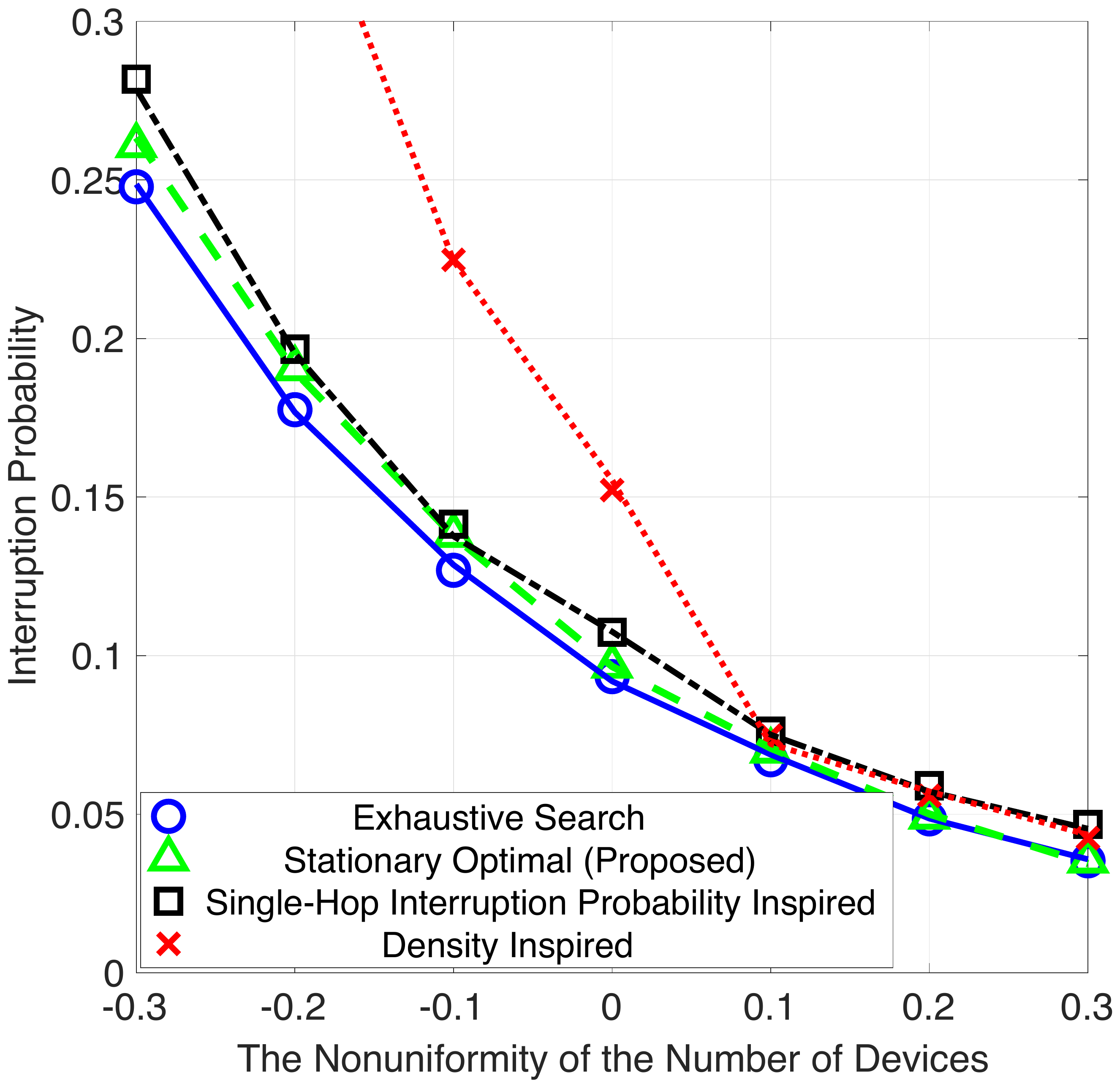}
\caption{Comparison of strategies under non-uniform device distributions.}
\label{fig:Figure7}
\end{minipage}
\hfill
\begin{minipage}[t]{0.48\linewidth}
\centering
\includegraphics[width=\linewidth]{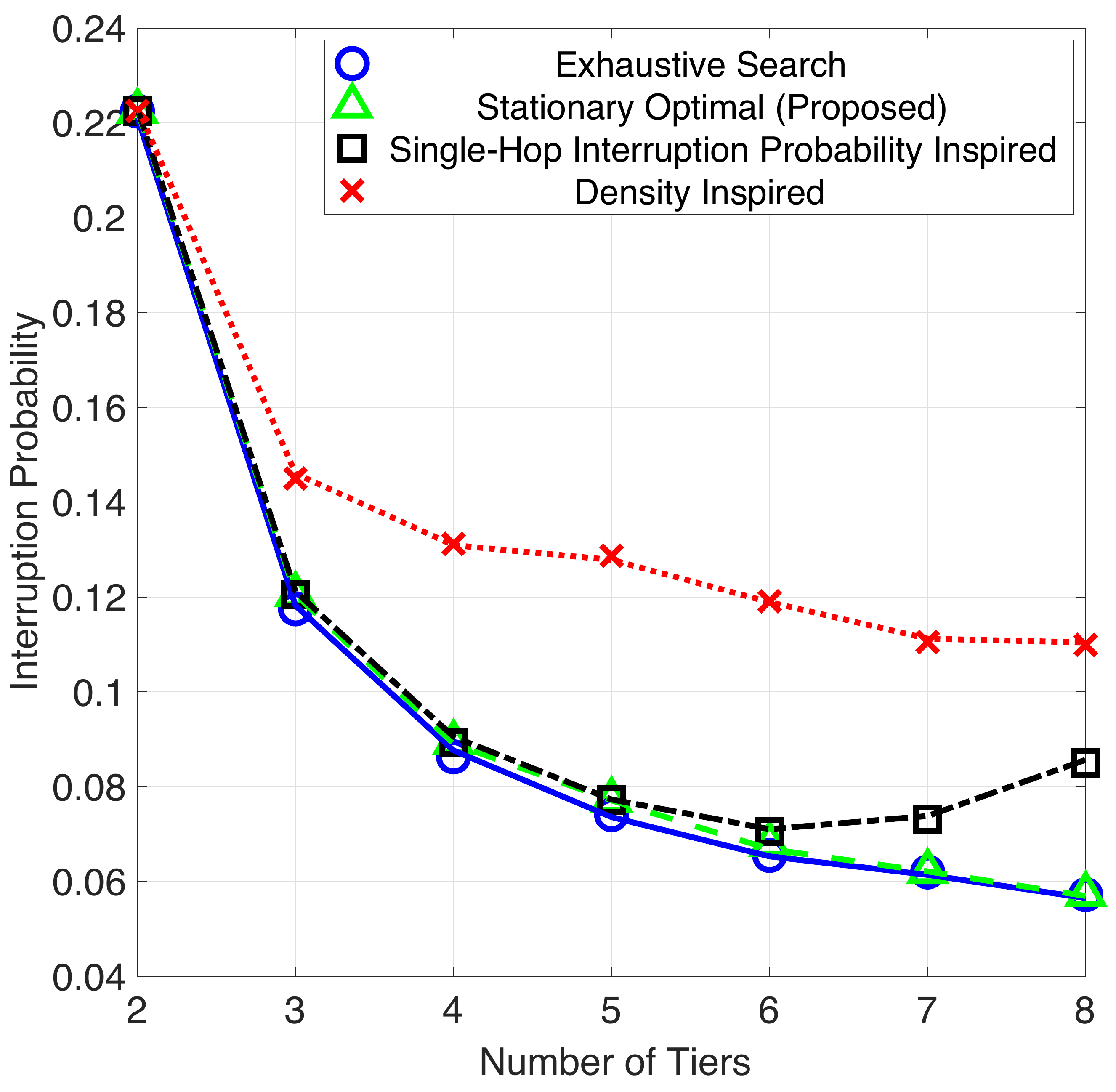}
\caption{Comparison of strategies under the different number of tiers.}
\label{fig:Figure12}
\end{minipage}
\end{figure*}

In both Fig.~\ref{fig:Figure7} and Fig.~\ref{fig:Figure12}, the total number of devices is fixed at 1500. In Fig.~\ref{fig:Figure7}, we consider a five-tier HSTRN consisting of $(1-2\alpha)\cdot 300$, $(1-\alpha)\cdot 300$, $300$, $(1+\alpha)\cdot 300$, $(1+2\alpha)\cdot 300$ relay devices with heights $0, 300, 600, 900$ and $1200$km, where $\alpha$ is the nonuniformity of the number of devices. In Fig.~\ref{fig:Figure12}, each tier contains the same number of relay devices, and the altitudes of the satellite tiers are equidistant points between $300$ km and $1200$ km. For example, in a three-tier HSTRN, there are one terrestrial tier and two satellite tiers. The altitudes of these two tiers are three equal points between $300$ km and $1200$ km, that is, $600$ km and $900$ km.

\par
As shown in Fig.~\ref{fig:Figure7}, the inverted pyramid relay device distribution ($\alpha > 0$) is more reliable than the pyramid ($\alpha < 0$). Both Fig.~\ref{fig:Figure7} and Fig.~\ref{fig:Figure12} show that the reliability performance of the stationary optimal strategy is close to the optimal one obtained by the exhaustive search. The performance of the single-hop interruption probability inspired strategy is only slightly inferior to the stationary optimal strategy. The density inspired strategy is a potential alternative only when relay devices are typically inverted pyramid distributed. Overall, there exists a balance between complexity and interruption probability. As shown in Fig.~\ref{fig:Figure12}, as the number of tiers increases, this single-hop interruption probability inspired strategy has an obvious effect on reducing complexity but also sacrificing reliability performance to a greater extent.

\subsection{Complementarity Between Terrestrial Devices and Satellites}\label{Deployment of Devices}
\begin{figure*}[htbp]
\begin{minipage}[t]{0.48\linewidth}
\centering
\includegraphics[width=\linewidth]{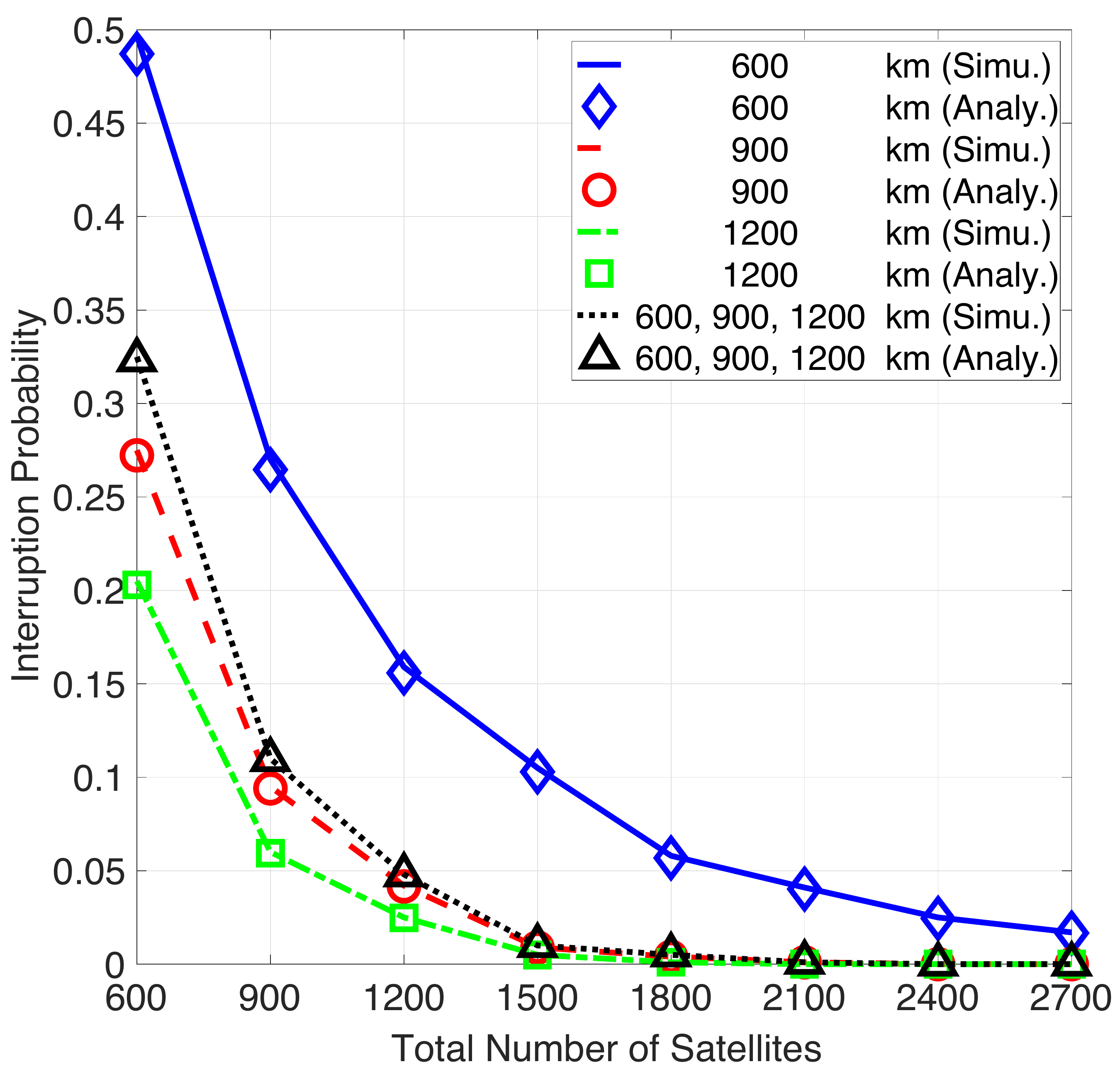}
\caption{The influence of the height and number of satellites on interruption probability.}
\label{fig:Figure6}
\end{minipage}
\hfill
\begin{minipage}[t]{0.48\linewidth}
\centering
\includegraphics[width=\linewidth]{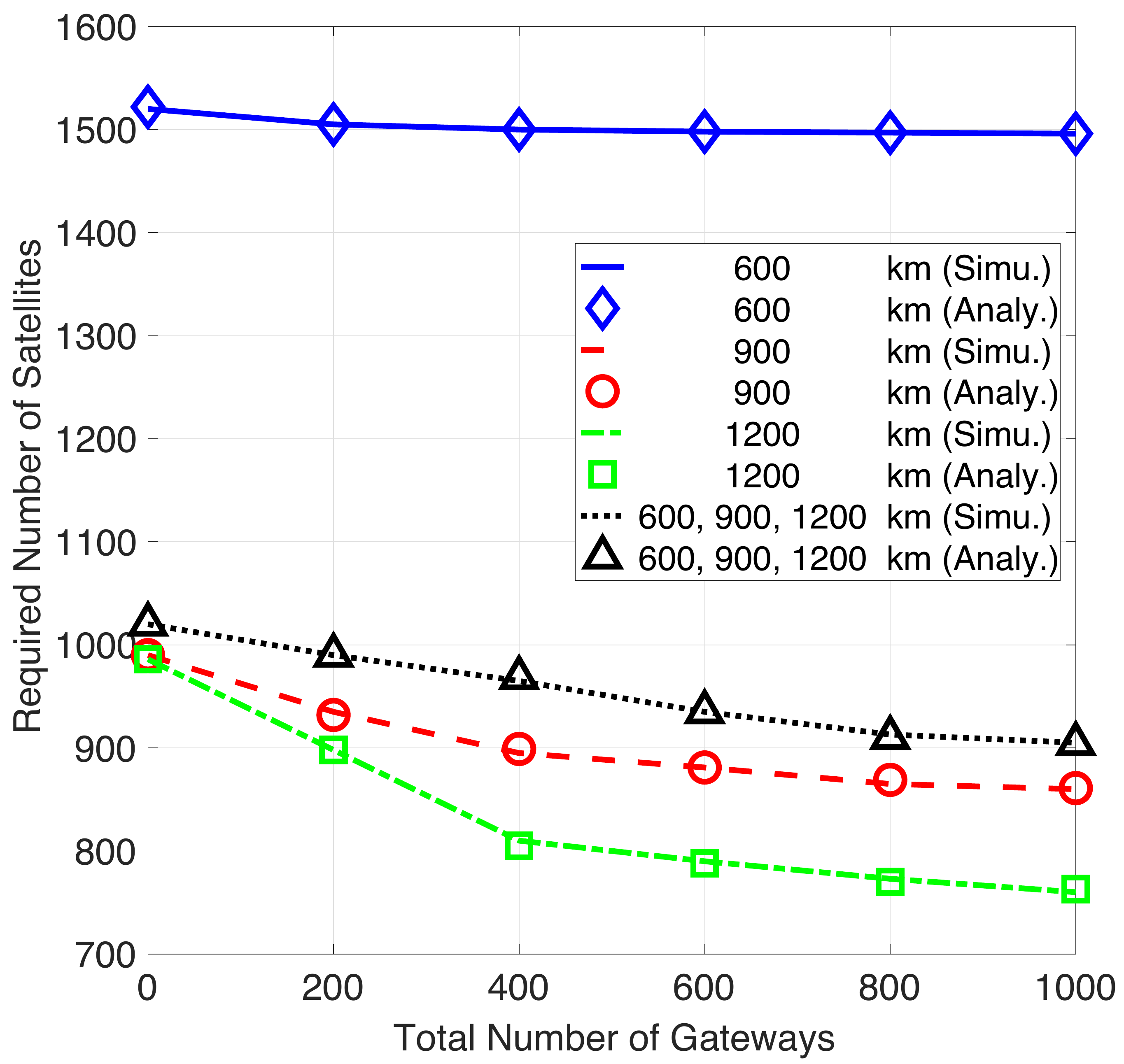}
\caption{The complementarity between gateways and satellites when interruption probability is $0.1$.}
\label{fig:Figure8}
\end{minipage}
\end{figure*}

One point about Fig.~\ref{fig:Figure12} remains to be explained: As the number of tiers increases, the interruption probability decreases. We propose two hypotheses: (i) Spreading satellites more uniformly across space (deploying more layers of satellites) reduces the interruption probability, and (ii) deploying more satellites rather than gateways reduces the interruption probability. Fig.~\ref{fig:Figure6} and Fig.~\ref{fig:Figure8} are used to verify these two hypotheses, respectively. In Fig.~\ref{fig:Figure6} and Fig.~\ref{fig:Figure8}, the black dotted line corresponds to a four-tier network including three satellite tiers at altitudes of $600$km, $900$km, and $1200$km with the same number of satellites. The stationary optimal priority strategy is adopted in this subsection. The total number of terrestrial gateways is fixed as $N_1 = 500$ in Fig.~\ref{fig:Figure6}, and the interruption probability is kept at 0.1 in Fig.~\ref{fig:Figure8}.

Fig.~\ref{fig:Figure6} proves that a three-tier network has no advantages over a single-tier network with respect to reliability. The performance of a three-tier network composed of satellites is close to a network composed of the same number of satellites at altitudes of $900$ km alone. Furthermore, the satellite with a higher altitude plays a more significant role in increasing reliability. Fig.~\ref{fig:Figure8}, shows that deploying more satellites is far more effective than gateways. Therefore, hypothesis (ii) is more suitable to explain the decreasing tendency in Fig.~\ref{fig:Figure12}. For a given target interruption probability, increasing the number of gateways can reduce the required number of satellites to different degrees. The effect of increased ground relay for the network with three tiers of satellites is roughly equal to the superposition of the respective roles of the three satellite networks.

\subsection{The Extension of Metrics Analysis}
In this subsection, we relax the assumptions in the system model and show the robustness of the scheme. In addition to reliability analysis, our framework is also applicable to the analysis of other widely studied metrics. We take satellite availability, coverage probability, and URLLC rate as examples. 

\par
\subsubsection{Satellite availability} 
The satellites of a link are said to be available if all the relay satellites on the link are within the joint line-of-sight region of the previous hop and the next hop. The probability that all satellites in the link are available is numerically equal to $1 - \widetilde{P}^M$ ($\widetilde{P}^M$ is defined in Theorem~\ref{theorem1}), as long as the following substitution is made for the maximum dome angle: 
\begin{equation}
    \theta_{i,j} = \max \left\{ \theta_{s},  \arccos\left(\frac{R_1}{R_i}\right)+\arccos\left(\frac{R_1}{R_j}\right) \right\},
\end{equation}
where $\theta_{i,j}$ the maximum dome angle of relays from tier $i$ and $j$.

\par
\subsubsection{Coverage probability} 
The coverage probability of a link is defined as the probability that the received SNR of each hop is greater than threshold $\gamma$. We assume that inter-satellite links follow the free space propagation model with no turbulence and perfect alignment, and satellite-terrestrial links follow a widely accepted channel fading model given in \cite{talgat2020stochastic}, the received SNR is given as,
\begin{equation}
\left\{\begin{matrix}
    &{\rm{SNR}}_1 = \frac{\rho G}{\sigma^2} \left( \frac{c}{4\pi f d} \right)^2 \zeta \mathcal{S} , & {\rm{for \ satellite-terrestrial \ links}}, \\
    &{\rm{SNR}}_2 = \frac{\rho G}{\sigma^2} \left( \frac{c}{4\pi f d} \right)^2, & {\rm{for \ inter-satellite \ links}},
\end{matrix}\right.
\end{equation}
where parameters' definitions and values are shown in Table~\ref{table3}. The coverage probability $P^C$ of the link can be approximately expressed as: 
\begin{equation}\label{PC}
    P^C \left(\gamma\right) \approx \mathbbm{P} \left[ {\rm{SNR}}_1 > \gamma \right]^2 \cdot \left( v_1 \mathbbm{P} \left[ {\rm{SNR}}_1 > \gamma \right] + \left(1-v_1\right) \cdot \mathbbm{P} \left[ {\rm{SNR}}_2 > \gamma \right] \right)^{N_h-2},
\end{equation}
where the analytical expression of $\mathbbm{P} \left[ {\rm{SNR}} > \gamma \right]$ can be obtained through derivation from Theorem~1 in \cite{talgat2020stochastic}.

\par
\subsubsection{URLLC rate}
We define the URLLC rate as the joint probability that the received SNR of each hop is greater than threshold $\gamma$ and the total latency is smaller than threshold $\tau$. The total latency is considered to be the propagation latency plus the buffering latency. The propagation latency is the time duration taken for a signal to travel through the air. Consider that signals travel at the speed of light in the form of electromagnetic waves, the propagation latency is calculated by $\sum_i \left ( \frac{d_i}{c} \right )$. The buffering latency is the time duration taken for a packet to be transmitted at the achievable data rate, and it can be calculated by $\sum_i \frac{\varpi }{B \log_{2}\left(1+{\rm{SNR}} \left(d_i\right)\right)}$, where $\varpi$ and $B$ denote package size and carrier bandwidth, respectively. The URLLC rate can be approximately expressed as: 
\begin{equation}
    P^U \left(\gamma,\tau \right) \approx P^C \left(\gamma\right) \cdot P^C \left( \exp \left( \frac{\tau \ln 2 }{N_h B} - \frac{\ln 2 \cdot \left(2\overline{d_1} + (N_h - 2 )\overline{d_2} \right) }{c N_h B} \right) - 1 \right),
\end{equation}
where $c=3\times10^8$m/s is the speed of light, $P^C \left(\gamma\right)$ is defined in (\ref{PC}), $\overline{d_1}$ and $\overline{d_2}$ are defined as \begin{equation}\label{d1d2}
\begin{split}
    \overline{d_1} & =  \sum_{i \in \mathcal{K}} v_i \sqrt{R_1^2+R_i^2-2R_1 R_i \cos\overline{\theta}_o} \, , \\
    \overline{d_2} = & \sum_{i \in \mathcal{K}} \sum_{j \in \mathcal{K}} v_i v_j \sqrt{R_i^2 + R_j^2- 2 R_i R_j \cos\overline{\theta}_o} \, ,
\end{split}
\end{equation}
where $\overline{\theta}_o$, $v_i$, and $N_h$ have the same definitions introduced in (\ref{theta_average}).

\par
\begin{figure*}[htbp]
\begin{minipage}[t]{0.47\linewidth}
\centering
\includegraphics[width=\linewidth]{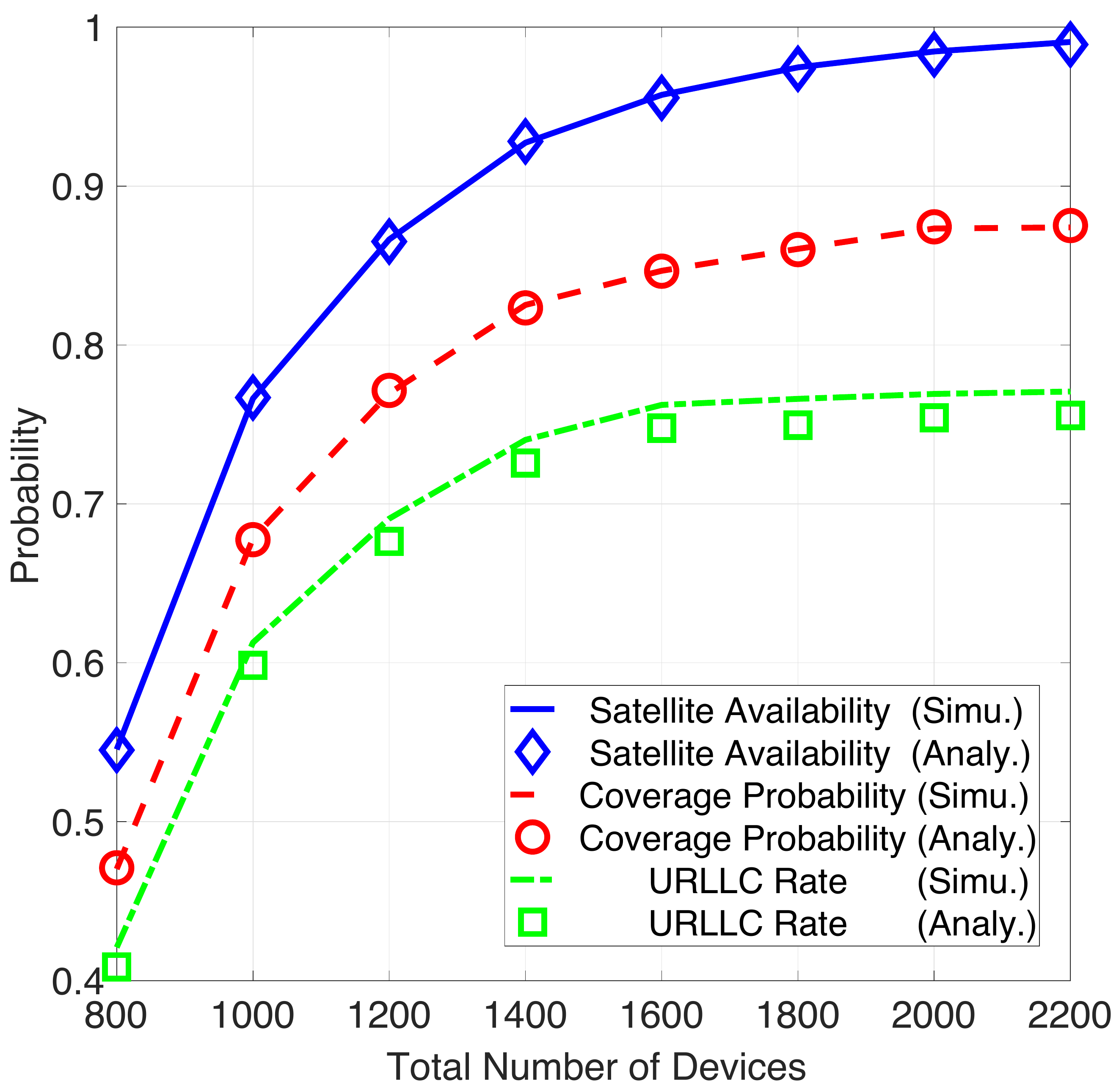}
\caption{The influence of the number of devices on metrics.}
\label{fig:Figure13}
\end{minipage}
\hfill
\begin{minipage}[t]{0.51\linewidth}
\centering
\includegraphics[width=\linewidth]{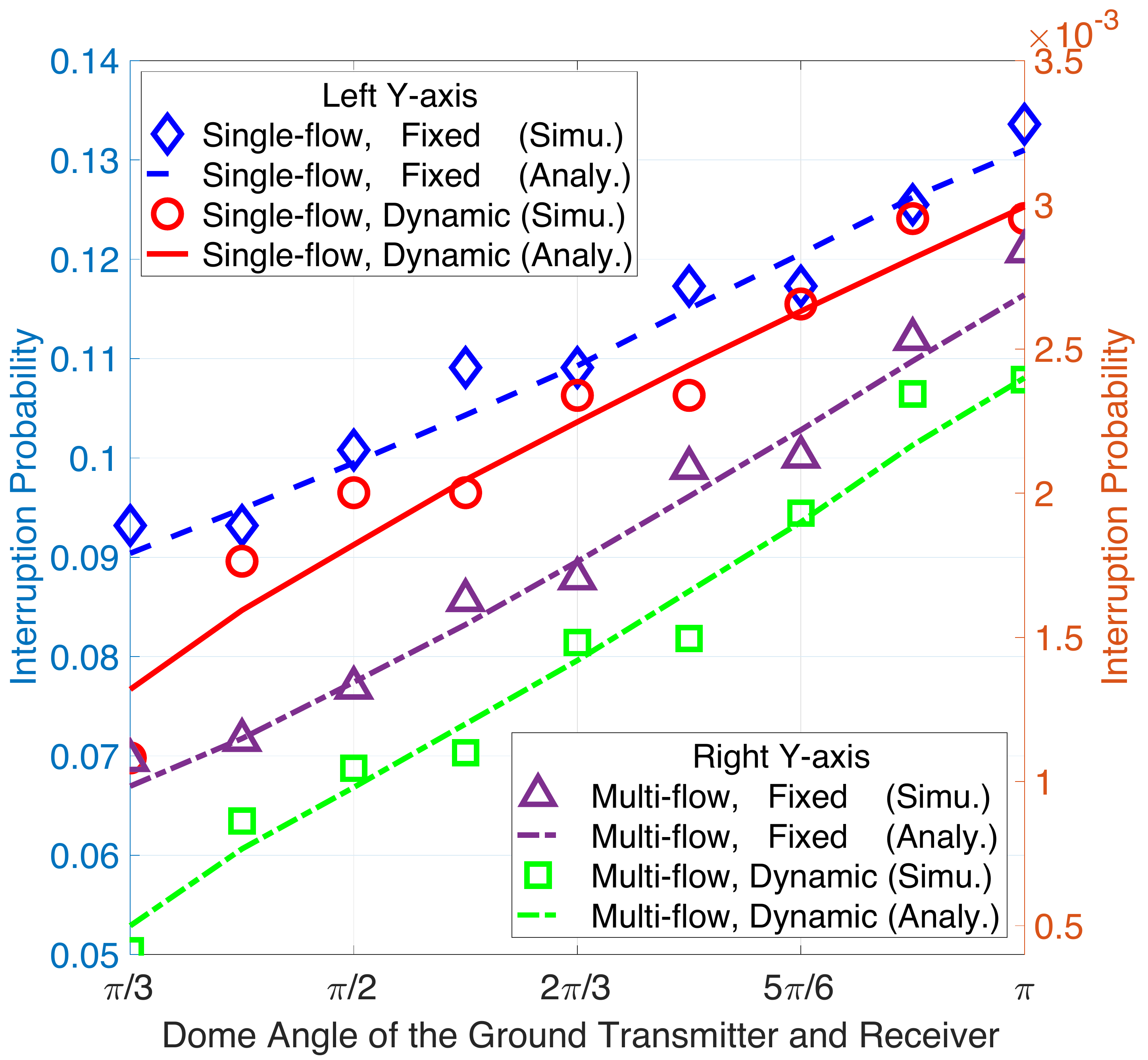}
\caption{Interruption probability for intelligent routing schemes.}
\label{fig:Figure11}
\end{minipage}
\end{figure*}

In Fig.~\ref{fig:Figure13}, we consider a four-tier network with heights of $0$, $600$, $900$, and $1200$~km with the same number of devices in each tier. The stationary optimal priority strategy is applied. The analytical results coincide well with the Monte Carlo simulation, proving that the reliability analytical framework in this article is also applicable to the analysis of other metrics. As the number of communication devices exceeds $1600$, further increasing devices has limited performance improvement for any of the three metrics.

\begin{table}[]
\centering
\caption{Value and definition of parameters \cite{talgat2020stochastic}.}
\label{table3}
\begin{tabular}{|c|c|c||c|c|c|}
\hline
Notation  & Parameter                & Value     & Notation & Parameter      & Value               \\ \hline  \hline
$f$     & Carrier frequency        & $20$~GHz  & $G$      & Total antenna gain & $41.7$~dBi \\ \hline
$\rho$ & Transmission power    & $ 15$~dBW & $B$    & Bandwidth      & $ 100$~MHz          \\ \hline
$\zeta$   & Average rain attenuation & $ -2$~dB  & $\varpi$ & Package size   & $100$~Mbits         \\ \hline
$\sigma ^2$ & Noise power & $3.6 \times 10^{-12}$~W                         & $\tau$         & Threshold of latency & $ 4 $~s \\ \hline
$\mathcal{S}$             & SR fading   & $ \mathcal{SR}(1.29,0.158,19.4)$ & $\gamma$ & Threshold of coverage probability   & $0$~dB      \\ \hline
\end{tabular}
\end{table}

\subsection{Intelligent Routing Schemes}
In this article, the link routes along the shortest inferior arc from the ground transmitter to the ground receiver. Furthermore, once the priority strategy is determined, it does not change during the routing (except for the penultimate hop). Therefore, we will discuss more intelligent routing schemes in this subsection.

\par
\subsubsection{Multi-flow routing scheme}
Except for the shortest inferior arc, other inferior arcs around the shortest inferior arc can also be regarded as potential routing flows. As long as any flow is not interrupted, the route is successful. Assume other inferior arcs are intersecting lines between the plane passing through the transmitter and receiver and the Earth or a sphere where satellites are located. Denote the dihedral angle between the corresponding planes of a chosen inferior arc and the shortest inferior arc as $\Theta$. Then, the interruption probability of this chosen flow (inferior arc) is approximately given by,
\begin{equation}
    \widehat{P}^M \left( \Theta \right) \approx \widetilde{P}^M \cdot \frac{\theta_m}{2 \sqrt{1-\left( \cos\frac{\theta_m}{2} \sin\Theta \right)^2 } } \cdot \frac{1}{\arcsin \frac{\sin\frac{\theta_m}{2}}{\sqrt{1-\left( \cos\frac{\theta_m}{2} \sin\Theta \right)^2 }}},
\end{equation}
where $\widetilde{P}^M=\widehat{P}^M \left( 0 \right)$ is defined in (\ref{p_I}) and $\theta_m$ is the dome angle between the ground transmitter and receiver. We take a multi-flow routing with three flows as an example. One flow corresponds to the shortest inferior arc, and the corresponding dihedral angle of the remaining two is $\pi/6$. Since these three flows are far apart, the probability of including common satellites in multi-flow is low, and the total interruption probability is
\begin{equation}
    \widehat{P}_{\rm{total}}^M = 1 - \left(1-\widehat{P}^M \left( \frac{\pi}{6} \right) \right)^2 \cdot \left( 1 - \widetilde{P}^M \right).
\end{equation}

\par
\subsubsection{Dynamic priority strategy}
The priority strategy of each hop needs to be updated based on the located tier and the remaining number of hops. This extension is not challenging because we can calculate multi-hop interrupt probabilities starting at different tiers by the expression provided in Theorem~\ref{theorem1}.

\par
Fig.~\ref{fig:Figure11} shows that the multi-flow routing scheme provides a much more significant improvement in reliability performance than the dynamic priority strategy. Note that the interruption probability for single-flow corresponds to the vertical axis on the left and for multi-flow corresponds to the right. The increase of interruption probability is almost linear with $\theta_m$. Because $N_h$ is evaluated discretely, the analysis results are dotted alternately above and below the simulation curve.



\section{Conclusion and Future Works}
This article provided a reliability analytical framework based on stochastic geometry for satellite networks. The special topology of satellite networks made our issues different from other routing reliability-related studies. We took multi-hop interruption probability as a measure of reliability and provided an analytic expression for interruption probability. The stationary optimal priority strategy that can approximate the minimum interruption probability and its low-complexity alternative was given. Furthermore, We studied how to achieve higher reliability through network deployment, and how to extend the interruption probability analysis to the analysis of other common network metrics.

\par
As for future works, stochastic geometry-based analysis results of routing coverage probability and URLLC rate have not yet been studied. Providing analytical frameworks for the mentioned intelligent routing schemes is also an interesting topic. If network traffic is considered, the congestion might affects the availability analysis of the network. In this case, satellite constellations can be modeled by Marked Poisson point processes (MPPPs) instead of BPPs. The locations of the satellites follow concentric spherical Poisson point processes, and the markers on the satellites record the degree of congestion. In particular, when the distribution of congestion degree satisfies the binomial distribution, the MPPPs can be regarded as thinning of Poisson point processes (PPPs). Furthermore, when analyzing the single-hop reliability, the pointing error caused by the relative movement of two satellites is also worth considering. In this case, a more accurate model, such as the orbit geometry model \cite{wang2022evaluating}, is required.

\appendices
\section{Proof of Lemma~\ref{lemma1}}\label{app:lemma1}
When $i \neq j$, the transmitter in tier $i$ does not belong to the point process in tier $j$. By definition, the interruption probability is equal to the probability that there are no available relays in the shadow region of Fig.~\ref{fig:Figure2}. Therefore,
\begin{equation}\label{Proof of contact angle}
\begin{split}
    P_{i,j}^{I} &= \mathbb{P}\left[ {\mathcal{N}\left( \mathcal{A} \right) = 0} \right] \overset{(a)}{=} \left( 1 - \frac{\mathcal{A}}{4 \pi R_j^2} \right)^{N_j}\\
    &\overset{(b)}{=} \left( 1 - \frac{1}{4 \pi R_j^2}\left( \frac{\theta_r }{2 \pi}2 \pi R_j \left( R_j-R_j\cos\theta_{i,j} \right)  - \frac{\theta_r }{2 \pi}2 \pi R_j \left( R_j-R_j\cos\theta_{s} \right) \right)\right)^{N_j} \\ 
    & = \left(1- \frac{\theta_r}{4\pi}\left(\cos\theta_s - \cos\theta_{i,j}\right) \right)^{N_j},
\end{split}
\end{equation}
where $\mathcal{A}$ is the area of the shadow region of Fig.~\ref{fig:Figure2}, and $\mathcal{N}\left( \mathcal{A} \right)$ counts the number of devices in $\mathcal{A}$. In step (a), for a homogeneous BPP, the probability that a single device is located in the shadow region is equal to the ratio of the area of the shadow region to the area of the entire sphere. Step (b) comes from the area formula of the spherical cap, where $R_j-R_j\cos\theta_{i,j}$ and $R_j-R_j\cos\theta_s$ are the height of the spherical caps. 
\par
When $i = j$, the transmitter belongs to the point process in tier $j$. Therefore, it searches the remaining $N_j-1$ devices for the next relay. As is shown in Fig.~\ref{fig:Figure2}, the above conclusion applies to both lower-to-upper perspective ($i<j$) and higher-to-lower ($i>j$) perspective.

\section{Proof of Lemma~\ref{mu_1}}\label{app:mu_1}
Firstly, $K$ states correspond to $K$ tiers and the absorbing state. The $K+1$ states can be further divided into positive recurrent, reducible, and transient states. The reducible states consist of unreachable tiers after many hops from the first tier. Since it does not affect the final analysis, we can omit the reducible states and only discuss the subsystem consisting of the remaining two states. It is easy to know that routing is bound to interrupt after infinite hops in such an irreducible subsystem. Therefore, except for the absorbing state representing the interruption is positive recurrent, the other states are all transient. The indexes of transient states are denoted as $\mathcal{K}$. To calculate the average number of hops before the interruption, the following theorem about the expected hitting times \cite{chen2008expected} is introduced. 

\begin{equation}
\mu_i = 
\left\{\begin{matrix}
1 + &\sum_{j=1}^{K+1} \widetilde{T}_{i,j}^{(2)} \mu_j,\ &i \in \mathcal{K}, \\ 
&0, &i \not\in \mathcal{K},
\end{matrix}\right.
\end{equation}
where $\mu_i$ represents the average number of hops interruption occurred after starting from tier $i$, $i \in \mathcal{K}$. There are some additions to the above conclusion. 
\begin{itemize}
    \item Since routing is assumed to start from the terrestrial tier, we only focus on $\mu_1$.
    \item In most cases, solving $\mu_1$ still requires solving the entire system of equations, where the number of equations is equal to the number of elements in set $\mathcal{K}$.
    \item In general, there is not much difference in value between different $\mu_i$.
\end{itemize}

\section{Proof of Lemma~\ref{lemma3}}\label{app:lemma3}
The proof of lemma~\ref{lemma3} is divided into two steps. Firstly, we prove that the average number of hops $N_h$ is obtained in the direction of the shortest path routing. Then, the expression of the average dome angle $\overline{\theta}_o$ is derived.
\par
Since there are infinite routing paths on the sphere from one point to another, the discussion begins with a path consisting of an arc from one circle between two points on the same sphere. Among all circles passing the transmitter and receiver on the sphere, the circle centered at the sphere's origin has the largest radius. Since these arcs have a common string that connects two points, the larger the radius, the shorter the arc length. Irregular paths can also be approximated as part of the arcs in small local regions. The above analysis method is similar for two points on different spheres. Therefore, the inferior arc corresponding to dome angle $\theta_m$ has the shortest length, and the direction following this arc is the direction of shortest path routing. 
\par
The following procedures are used to derive the analytic expression for the average dome angle $\overline{\theta}_o$. Suppose a hop starts at tier $i$ and ends at tier $j$. The average area searched for the nearest relay device near an arbitrary reference point can be expressed by the following formula. Since the azimuth angle of BPP is uniformly distributed, we use the spherical cap area to express the average search area $\overline{\mathcal{A}}_{i,j}$,
\begin{equation}\label{A1}
    \overline{\mathcal{A}}_{i,j} = 2 \pi R_j^2 \left( 1 - \cos\mathbb{E}\left[\theta_{\rm{cap},i,j}\right] \right),
\end{equation}
where $\theta_{\rm{cap},i,j}$ is the dome angle of the spherical cap. When $i=j$, $\theta_{\rm{cap},i,j}$ obeys nearest neighbor angle distribution. When $i \neq j$, $\theta_{\rm{cap},i,j}$ obeys contact angle distribution. When $i \neq j$, 

\begin{equation}
\begin{split}
\mathbb{E}\left[\theta_{\rm{cap},i,j}\right] &\approx \int_0^{\pi} 1 - F_{\theta_{\rm{cap},i,j}}\left(\theta\right)\mathrm{d}\theta  = \int_0^{\pi} 1 - \left( 1 - \mathbb{P}\left[ {\mathcal{N}\left( \overline{\mathcal{A}}_{i,j} \right) = 0} \right] \right)
\mathrm{d}\theta \\
& = \int_0^{\pi} \left( 1 - \frac{2 \pi R_j \left( R_j-R_j\cos\theta \right)}{4 \pi R_j^2} \right)^{N_j} \mathrm{d}\theta  =  \int_0^{\pi} \left(\frac{1+\cos\theta}{2}\right)^{N_j}
\mathrm{d}\theta\\
& = 2\int_0^{\frac{\pi}{2}}\left(\cos\theta \right)^{2N_j} \mathrm{d}\theta \overset{(a)} = \pi \prod \limits_{k=1}^{N_j} \frac{2k-1}{2k},
\end{split}
\end{equation}
where (a) follows Wallis' integrals \cite{dana2012parametric}, $\mathcal{N}\left( \overline{\mathcal{A}}_{i,j} \right)$ counts the number of devices in the spherical cap $\mathcal{A}$. Since some part of the proof is similar to that of (\ref{Proof of contact angle}), therefore omit it here. Since the device closest to the receiver is preferentially selected as the relay, so the area between the average dome angle $\overline{\theta}_{o,i,j}$ and the maximum dome angle $\theta_{i,j}$ should also be equal to the average search area,
\begin{equation}\label{A2}
    \overline{\mathcal{A}}_{i,j} = \theta_r R_j^2 \left(\cos\overline{\theta}_{o,i,j} - \cos\theta_{i,j} \right).
\end{equation}
The above formula is obtained by substituting $\overline{\theta}_{o,i,j}$ to $\theta_s$ in (\ref{Proof of contact angle}). Notice that the area between $\overline{\theta}_{o,i,j}$ and $\theta_{i,j}$ is part of the circular ring that corresponds to the central angle $\theta_r$. The shape is similar to the shadow area in Fig.~\ref{fig:Figure2}. Simultaneous (\ref{A1}) and (\ref{A2}), we have
\begin{equation}\label{theta_o,i,j}
    \overline{\theta}_{o,i,j} = \arccos\left( \frac{2 \pi}{\theta_r} - \frac{2 \pi}{\theta_r} \cos\left(\pi \prod_{k=1}^{N_j}\frac{2k-1}{2k} \right) + \cos\theta_{i,j} \right).
\end{equation}
Similarly, when $i = j$, subtract one from $N_j$ in (\ref{theta_o,i,j}),
\begin{equation}\label{theta_o,i,i}
    \overline{\theta}_{o,i,i} = \arccos\left(  \frac{2 \pi}{\theta_r} - \frac{2 \pi}{\theta_r} \cos\left(\pi \prod_{k=1}^{N_i-1}\frac{2k-1}{2k} \right) + \cos\theta_{i,i} \right).
\end{equation}

\par
Finally, $\overline{\theta}_{o}$ is obtained by taking a weighted average of $\overline{\theta}_{o,i,j}$. $\overline{\theta}_{o,i,j}$ needs to be weighted by the product of the probability of the hop starting at tier $i$ and the probability that the hop ends at tier $j$,
\begin{equation}
\begin{split}\label{theta_o,i,j and i,i}
    \overline{\theta}_o = \sum_{i=1}^{K} v_i \sum_{j=1,j \neq i}^{K} T_{i,j}^{(1)} \overline{\theta}_{o,i,j} + \sum_{i=1}^{K} v_i T_{i,i}^{(1)} \overline{\theta}_{o,i,i}.
\end{split}
\end{equation}
Substitute (\ref{theta_o,i,j}) and (\ref{theta_o,i,i}) into (\ref{theta_o,i,j and i,i}), the lemma is proved.

\section{Proof of Proposition~\ref{proposition2}}\label{app:proposition2}
By analyzing the process of solving the interruption probability, it is easy to know that only the augmented matrices $\widetilde{T}^{(2)}$ and $\widehat{T}^{(3)}$ may not be unique. Since operators $\widetilde{\mathcal{T}}^{(2)}$ and  $\widehat{\mathcal{T}}^{(3)}$ only contain basic subtraction and multiplication operations, and $P^I$ is unique, only the uniqueness of $s^*$ is uncertain. Therefore, the following three cases are discussed to prove the uniqueness of the multi-hop interruption probability. 
\begin{itemize}
    \item When all of the tiers are irreducible, the stationary distribution is unique. Under this condition, $s^*$ and the multi-hop interruption probability $p_I$ are also unique.
    \item If the first tier is reducible and zero recurrence, it indicates that communication is bound to fail and the interruption probability $p_I=0$, which is unique.
    \item If tier $i$ ($i \neq 1$) is reducible and zero recurrence, $\widetilde{\mathcal{T}}^{(2)}$ and  $\widehat{\mathcal{T}}^{(3)}$ might have different values of elements at $i^{th}$ row and column. However, since tier $i$ is unreachable when starting from the first tier, no matter how many hops it goes through, the $i^{th}$ element of row vectors $e_1 \widetilde{T}^{(2)}, \, e_1 \left(\widetilde{T}^{(2)}\right)^2 \, , \dots, e_1 \left(\widetilde{T}^{(2)}\right)^{N_h-2} \widehat{T}^{(3)}$ are always 0. Therefore, the interruption probability is unique, although $\widetilde{\mathcal{T}}^{(2)}$ and $\widehat{\mathcal{T}}^{(3)}$ might have several expressions.
\end{itemize}

\section{Proof of Theorem~\ref{theorem1}}\label{app:theorem1}
In (\ref{p_I}), $e_1$ represents the probability of appearing in the first tier is 1, the probabilities of appearing in other tiers are 0, and the initial value of the interruption probability is also 0. The process of right-multiplying matrix $\widetilde{T}^{(2)}$ $n$ times means that $n$ hops have been completed. The first $K$ elements in the row vector $e_1 \left(\widetilde{T}^{(2)}\right)^n$ represent the probability of the relay occurring at each tier, and the last element represents the interruption probability after $n$ hops. 
\par
Since the routing contains $N_h$ hops in average, the last two hops are presented by right-multiplying $\widehat{T}^{(3)}$, and the first $N_h-2$ hops are done by right-multiplying $\widetilde{T}^{(2)}$ $N_h-2$ times. Finally, the last element of $e_1 \left(\widetilde{T}^{(2)}\right)^{N_h-2} \widehat{T}^{(3)}$, which is the interruption probability after $N_h$ hops, is extracted by $e_{K+1}^T$.

\section{Proof of Proposition~\ref{proposition1}}\label{app:proposition1}
When all tiers are reachable, the matrix $T^{(1)}$ is recurrent and irreducible, so it has a unique stationary distribution $v$. When $T^{(1)}$ is reducible, the submatrix, after removing the rows and columns corresponding to the unreachable tiers in $T^{(1)}$ is also irreducible. Therefore, the first part of the proposition has been proved. 
\par

As for the second part, according to the equation given in (\ref{weighted single-hop}),
\begin{equation}
    \overline{P}^S = \sum_{i=1}^{K} v_i P_{i}^S,
\end{equation}
we only need to prove the uniqueness of $P_{i}^S$ for all $i \leq K$, under the premise that the stationary distribution $v$ is given. From (\ref{single-hop interruption}), we know that $P_{i}^S$ is only determined by the communication technique, i.e. constraints $\left(c_1\right)$, $\left(c_2\right)$ and $\left(c_3\right)$, and the number of relay devices. 
\par

Furthermore,  note that the terms in (\ref{single-hop interruption}) can exist in order, for example, putting the $j=2$ term in the first place, which means that the second tier's relay satellites are searched first. Different multiplication orders correspond to different strategies. We can prove that the value of $P_{i}^S$ is independent of priority strategies in terms of the commutativity of multiplication. Therefore, the proposition is proved.

\bibliographystyle{IEEEtran}
\bibliography{references}

\begin{thebibliography}{10}
\providecommand{\url}[1]{#1}
\csname url@samestyle\endcsname
\providecommand{\newblock}{\relax}
\providecommand{\bibinfo}[2]{#2}
\providecommand{\BIBentrySTDinterwordspacing}{\spaceskip=0pt\relax}
\providecommand{\BIBentryALTinterwordstretchfactor}{4}
\providecommand{\BIBentryALTinterwordspacing}{\spaceskip=\fontdimen2\font plus
\BIBentryALTinterwordstretchfactor\fontdimen3\font minus
  \fontdimen4\font\relax}
\providecommand{\BIBforeignlanguage}[2]{{%
\expandafter\ifx\csname l@#1\endcsname\relax
\typeout{** WARNING: IEEEtran.bst: No hyphenation pattern has been}%
\typeout{** loaded for the language `#1'. Using the pattern for}%
\typeout{** the default language instead.}%
\else
\language=\csname l@#1\endcsname
\fi
#2}}
\providecommand{\BIBdecl}{\relax}
\BIBdecl

\bibitem{yue2022security}
P.~Yue, J.~An, J.~Zhang, G.~Pan, S.~Wang, P.~Xiao, and L.~Hanzo, ``On the
  security of {LEO} satellite communication systems: Vulnerabilities,
  countermeasures, and future trends,'' \emph{\rm{available online:}
  https://arxiv.org/abs/2201.03063}.

\bibitem{yaacoub2020key}
E.~Yaacoub and M.-S. Alouini, ``A key {6G} challenge and opportunity-connecting
  the base of the pyramid: {A} survey on rural connectivity,''
  \emph{Proceedings of the IEEE}, vol. 108, no.~4, pp. 533--582, 2020.

\bibitem{9568932}
O.~B. Osoro and E.~J. Oughton, ``A techno-economic framework for satellite
  networks applied to low earth orbit constellations: Assessing {S}tarlink,
  {O}ne{W}eb and {K}uiper,'' \emph{IEEE Access}, vol.~9, pp.
  141\,611--141\,625, 2021.

\bibitem{kodheli2020satellite}
O.~Kodheli, E.~Lagunas, N.~Maturo, S.~K. Sharma, B.~Shankar, J.~F.~M. Montoya,
  J.~C.~M. Duncan, D.~Spano, S.~Chatzinotas, S.~Kisseleff, J.~Querol, L.~Lei,
  T.~X. Vu, and G.~Goussetis, ``Satellite communications in the new space era:
  {A} survey and future challenges,'' \emph{IEEE Communications Surveys
  Tutorials}, vol.~23, no.~1, pp. 70--109, 2021.

\bibitem{chaudhry2020free}
A.~U. Chaudhry and H.~Yanikomeroglu, ``Free space optics for next-generation
  satellite networks,'' \emph{IEEE Consumer Electronics Magazine}, 2020.

\bibitem{zhu2021integrated}
X.~Zhu and C.~Jiang, ``Integrated satellite-terrestrial networks toward {6G}:
  Architectures, applications, and challenges,'' \emph{IEEE Internet of Things
  Journal}, vol.~9, no.~1, pp. 437--461, 2022.

\bibitem{zhang2021stochastic}
X.~Zhang, B.~Zhang, K.~An, G.~Zheng, S.~Chatzinotas, and D.~Guo, ``Stochastic
  geometry-based analysis of cache-enabled hybrid satellite-aerial-terrestrial
  networks with non-orthogonal multiple access,'' \emph{IEEE Transactions on
  Wireless Communications}, vol.~21, no.~2, pp. 1272--1287, 2022.

\bibitem{wang2022stochastic}
R.~Wang, M.~A. Kishk, and M.-S. Alouini, ``Stochastic geometry-based low
  latency routing in massive {LEO} satellite networks,'' \emph{IEEE
  Transactions on Aerospace and Electronic Systems}, to appear.

\bibitem{Al-3}
A.~Al-Hourani, ``A tractable approach for predicting pass duration in dense
  satellite networks,'' \emph{IEEE Communications Letters}, vol.~25, no.~8, pp.
  2698--2702, 2021.

\bibitem{tang2018multipath}
F.~Tang, H.~Zhang, and L.~T. Yang, ``Multipath cooperative routing with
  efficient acknowledgement for {LEO} satellite networks,'' \emph{IEEE
  Transactions on Mobile Computing}, vol.~18, no.~1, pp. 179--192, 2018.

\bibitem{petit2021assessment}
A.~Petit, A.~Rossi, and E.~M. Alessi, ``Assessment of the close approach
  frequency and collision probability for satellites in different
  configurations of large constellations,'' \emph{Advances in Space Research},
  vol.~67, no.~12, pp. 4177--4192, 2021.

\bibitem{lou2021green}
Z.~Lou, A.~Elzanaty, and M.-S. Alouini, ``Green tethered uavs for {EMF}-aware
  cellular networks,'' \emph{IEEE Transactions on Green Communications and
  Networking}, vol.~5, no.~4, pp. 1697--1711, 2021.

\bibitem{shen2020dynamic_2}
L.~Shen, Y.~Wang, L.~Liu, S.~Liu, D.~Wang, Y.~Fan, H.~Zhou, and T.~Ling, ``A
  dynamic modified routing strategy based on load balancing in {LEO} satellite
  network,'' in \emph{International Conference on Wireless and Satellite
  Systems}.\hskip 1em plus 0.5em minus 0.4em\relax Springer, 2020, pp.
  233--244.

\bibitem{geng2021agent_3}
S.~Geng, S.~Liu, Z.~Fang, and S.~Gao, ``An agent-based clustering framework for
  reliable satellite networks,'' \emph{Reliability Engineering \& System
  Safety}, vol. 212, p. 107630, 2021.

\bibitem{9348676_1}
J.~Hu, L.~Cai, C.~Zhao, and J.~Pan, ``Directed percolation routing for
  ultra-reliable and low-latency services in low earth orbit ({LEO}) satellite
  networks,'' in \emph{IEEE 92nd Vehicular Technology Conference
  (VTC2020-Fall)}, 2020, pp. 1--6.

\bibitem{knight2011internet}
S.~Knight, H.~X. Nguyen, N.~Falkner, R.~Bowden, and M.~Roughan, ``The internet
  topology zoo,'' \emph{IEEE Journal on Selected Areas in Communications},
  vol.~29, no.~9, pp. 1765--1775, 2011.

\bibitem{zhao2021multi_4}
N.~Zhao, X.~Long, and J.~Wang, ``A multi-constraint optimal routing algorithm
  in {LEO} satellite networks,'' \emph{Wireless Networks}, pp. 1--12, 2021.

\bibitem{8068282_5}
Y.~Cao, Y.~Shi, J.~Liu, and N.~Kato, ``Optimal satellite gateway placement in
  space-ground integrated network for latency minimization with reliability
  guarantee,'' \emph{IEEE Wireless Communications Letters}, vol.~7, no.~2, pp.
  174--177, 2018.

\bibitem{wang2022ultra}
R.~Wang, M.~A. Kishk, and M.-S. Alouini, ``Ultra-dense {LEO} satellite-based
  communication systems: {A} novel modeling technique,'' \emph{Communications
  Magazine}, vol.~60, no.~4, pp. 25--31, 2022.

\bibitem{dhillon2015wireless}
H.~S. Dhillon and G.~Caire, ``Wireless backhaul networks: {C}apacity bound,
  scalability analysis and design guidelines,'' \emph{IEEE Transactions on
  Wireless Communications}, vol.~14, no.~11, pp. 6043--6056, 2015.

\bibitem{farooq2015stochastic}
M.~J. Farooq, H.~ElSawy, and M.-S. Alouini, ``A stochastic geometry model for
  multi-hop highway vehicular communication,'' \emph{IEEE Transactions on
  Wireless Communications}, vol.~15, no.~3, pp. 2276--2291, 2015.

\bibitem{sasaki2017energy}
S.~Sasaki, Y.~Miyaji, and H.~Uehara, ``Energy budget formulation in
  progress-based nearest forwarding routing policy for energy-efficient
  wireless sensor networks,'' \emph{IEICE Transactions on Information and
  Systems}, vol. 100, no.~12, pp. 2808--2817, 2017.

\bibitem{routingimportant}
M.~Haenggi, ``On routing in random {R}ayleigh fading networks,'' \emph{IEEE
  Transactions on Wireless Communications}, vol.~4, no.~4, pp. 1553--1562,
  2005.

\bibitem{richter2018optimal}
Y.~Richter and I.~Bergel, ``Optimal and suboptimal routing based on partial
  {CSI} in random ad-hoc networks,'' \emph{IEEE Transactions on Wireless
  Communications}, vol.~17, no.~4, pp. 2815--2826, 2018.

\bibitem{wang2022evaluating}
R.~Wang, M.~A. Kishk, and M.-S. Alouini, ``Evaluating the accuracy of
  stochastic geometry based models for leo satellite networks analysis,''
  \emph{IEEE Communications Letters}, to appear.

\bibitem{Al-1}
A.~Al-Hourani, ``An analytic approach for modeling the coverage performance of
  dense satellite networks,'' \emph{IEEE Wireless Communications Letters},
  vol.~10, no.~4, pp. 897--901, 2021.

\bibitem{Al-2}
------, ``Optimal satellite constellation altitude for maximal coverage,''
  \emph{IEEE Wireless Communications Letters}, vol.~10, no.~7, pp. 1444--1448,
  2021.

\bibitem{talgat2020nearest}
A.~Talgat, M.~A. Kishk, and M.-S. Alouini, ``Nearest neighbor and contact
  distance distribution for binomial point process on spherical surfaces,''
  \emph{IEEE Communications Letters}, vol.~24, no.~12, pp. 2659--2663, 2020.

\bibitem{talgat2020stochastic}
------, ``Stochastic geometry-based analysis of leo satellite communication
  systems,'' \emph{IEEE Communications Letters}, vol.~25, no.~8, pp.
  2458--2462, 2021.

\bibitem{ok-1}
N.~Okati, T.~Riihonen, D.~Korpi, I.~Angervuori, and R.~Wichman, ``Downlink
  coverage and rate analysis of low {E}arth orbit satellite constellations
  using stochastic geometry,'' \emph{IEEE Transactions on Communications},
  vol.~68, no.~8, pp. 5120--5134, 2020.

\bibitem{ok-2}
N.~Okati and T.~Riihonen, ``Modeling and analysis of {LEO} mega-constellations
  as nonhomogeneous {P}oisson point processes,'' in \emph{IEEE 93rd Vehicular
  Technology Conference (VTC2021-Spring)}, 2021, pp. 1--5.

\bibitem{al2021modeling}
B.~Al~Homssi and A.~Al-Hourani, ``Modeling uplink coverage performance in
  hybrid satellite-terrestrial networks,'' \emph{Communications Letters},
  vol.~25, no.~10, pp. 3239--3243, 2021.

\bibitem{wang2022conditional}
R.~Wang, A.~Talgat, M.~A. Kishk, and M.-S. Alouini, ``Conditional contact angle
  distribution in {LEO} satellite-relayed transmission,'' \emph{IEEE
  Communications Letters}, to appear.

\bibitem{chen2008expected}
H.~Chen and F.~Zhang, ``The expected hitting times for finite {M}arkov
  chains,'' \emph{Linear Algebra and its Applications}, vol. 428, no. 11-12,
  pp. 2730--2749, 2008.

\bibitem{dana2012parametric}
T.~Dana-Picard and D.~G. Zeitoun, ``Parametric improper integrals, {W}allis
  formula and {C}atalan numbers,'' \emph{International Journal of Mathematical
  Education in Science and Technology}, vol.~43, no.~4, pp. 515--520, 2012.

\end{thebibliography}

\end{document}